\documentclass{emulateapj}
\usepackage{lscape}

\newcommand{\lbol}{\mbox{$L_{\rm bol}$}} % bolometric luminosity
\newcommand{\lint}{\mbox{$L_{\rm int}$}} % internal luminosity
\newcommand{\lmir}{\mbox{$L_{\rm MIR}$}} % mid-infrared luminosity
\newcommand{\tbol}{\mbox{$T_{\rm bol}$}} % bolometric temperature
\newcommand{\tbolprime}{\mbox{$T^{\prime}_{\rm bol}$}} % Tbol after ext. corr.
\newcommand{\lbolprime}{\mbox{$L^{\prime}_{\rm bol}$}} % Lbol after ext. corr.
\newcommand{\aprime}{\mbox{$\alpha^{\prime}$}} % alpha after ext. corr.
\newcommand{\andre}{Andr\'{e}}
\newcommand{\merin}{Mer\'{i}n}

\newcommand{\comeron}{Comer\'{o}n}
\newcommand{\jorgensen}{J{\o}rgensen}
\newcommand{\alcala}{Alcal{\'a}}

\newcommand{\degree}{\mbox{$^{\circ}$}}

\newcommand{\um}{$\mu$m}
\newcommand{\lsun}{\mbox{L$_\odot$}}% Lsun
\newcommand{\msun}{\mbox{M$_\odot$}}% Msun
\newcommand{\rsun}{\mbox{R$_\odot$}}% Msun

\slugcomment{\footnotesize Accepted for Publication in AJ}

\begin{document}
%%%%%%%%%%%%%%%%%% title %%%%%%%%%%%%%%%%%%%%%%%%%%%%%%%%%%%%%%%%
\title {\bf The Luminosities of Protostars in the \emph{Spitzer} c2d and 
Gould Belt Legacy Clouds}
\author{
Michael M.~Dunham\altaffilmark{1,2}, 
H\'ector G.~Arce\altaffilmark{1}, 
Lori E.~Allen\altaffilmark{3}, 
Neal J.~Evans II\altaffilmark{4}, 
Hannah Broekhoven-Fiene\altaffilmark{5}, 
Nicholas L.~Chapman\altaffilmark{6}, 
Lucas A.~Cieza\altaffilmark{7}, 
Robert A.~Gutermuth\altaffilmark{8}, 
Paul M.~Harvey\altaffilmark{4}, 
Jennifer Hatchell\altaffilmark{9}, 
Tracy L.~Huard\altaffilmark{10}, 
Jason M.~Kirk\altaffilmark{11}, 
Brenda C.~Matthews\altaffilmark{5}, 
Bruno Mer\'{i}n\altaffilmark{12}, 
Jennifer F.~Miller\altaffilmark{10, 13}, 
Dawn E.~Peterson\altaffilmark{14}, 
\& Loredana Spezzi\altaffilmark{15}
}

\altaffiltext{1}{Department of Astronomy, Yale University, P.O. Box 208101, New Haven, CT 06520, USA}

\altaffiltext{2}{michael.dunham@yale.edu}

\altaffiltext{3}{National Optical Astronomy Observatories, Tucson, AZ, USA}

\altaffiltext{4}{Department of Astronomy, The University of Texas at Austin, 2515 Speedway, Stop C1400, Austin, TX 78712-1205, USA}

\altaffiltext{5}{Herzberg Institute, National Research Council of Canada, 5071 W. Saanich Road, Victoria, BC V9E 2E7, Canada}

\altaffiltext{6}{Center for Interdisciplinary Exploration and Research in Astrophysics (CIERA), Department of Physics \& Astronomy, 2145 Sheridan Road, Evanston, IL 60208, USA}

\altaffiltext{7}{Institute for Astronomy, University of Hawaii at Manoa, Honolulu, HI 96822, USA}

\altaffiltext{8}{Department of Astronomy, University of Massachusetts, Amherst, MA, USA}

\altaffiltext{9}{Astrophysics Group, Physics, University of Exeter, Exeter EX4 4QL, UK}

\altaffiltext{10}{Department of Astronomy, University of Maryland, College Park, MD 20742, USA}

\altaffiltext{11}{School of Physics and Astronomy, Cardiff University, Queens Buildings, The Parade, Cardiff, CF24 3AA, UK}

\altaffiltext{12}{Herschel Science Centre, ESAC-ESA, P.O. Box 78, 28691 Villanueva de la Ca\~nada, Madrid, Spain}

\altaffiltext{13}{Harvard–Smithsonian Center for Astrophysics, 60 Garden Street, Cambridge, MA 02138, USA}

\altaffiltext{14}{Space Science Institute, 4750 Walnut Street, Suite 205, Boulder, CO 80301}

\altaffiltext{15}{European Southern Observatory (ESO), Karl-Schwarzschild-Strasse 2, D-85748 Garching bei M\"{u}nchen, Germany}

\begin{abstract}
Motivated by the long-standing ``luminosity problem'' in low-mass star 
formation whereby protostars are underluminous compared to theoretical 
expectations, we identify 230 protostars in 18 molecular clouds observed by two 
\emph{Spitzer Space Telescope} Legacy surveys of nearby star-forming regions.  
We compile complete spectral energy distributions, 
calculate \lbol\ for each source, and study the protostellar luminosity 
distribution.  This distribution extends over three orders of 
magnitude, from 0.01 \lsun\ -- 69 \lsun, and has a mean and median of 4.3 
\lsun\ and 1.3 \lsun, respectively.  The distributions are very 
similar for Class 0 and Class I sources except for an excess of low luminosity 
(\lbol\ $\la 0.5$ \lsun) Class I sources compared to Class 0.  100 out of the 
230 protostars (43\%) lack any available data in the far-infrared 
and submillimeter (70 \um\ $< \lambda <$ 850 \um) 
and have \lbol\ underestimated by factors of 2.5 on average, 
and up to factors of $8-10$ in extreme cases.  Correcting these 
underestimates for each source individually once additional data becomes 
available will likely increase both the mean and median of the sample 
by 35\% -- 40\%.  We discuss and compare our results to several recent 
theoretical studies of protostellar luminosities and show that our new results 
do not invalidate the conclusions of any of these studies.  As these studies 
demonstrate that there is more than one plausible accretion scenario that 
can match observations, future attention is clearly needed.  
The better statistics provided by our increased dataset should aid such 
future work.
\end{abstract}


%%%%%%%%%%%%%%%%%%%%%%%%%%%%%%%%%%%%%%%%%%%%%%%%%%%%%%%%%%%%%

\section{Introduction}\label{sec_intro}

Low-mass stars form from the gravitational collapse of dense molecular 
cloud cores of gas and dust (e.g., Beichman et al.~1986; Di Francesco et 
al.~2007).  During the collapse process material accretes from the core 
onto the protostar.  In this paper the term protostar is used to refer 
to the hydrostatic object at the center of a collapsing core.  More 
evolved young stellar objects (YSOs) no longer embedded within and forming 
from their natal dense cores are not considered protostars by this 
definition.

Despite several decades of progress, many details relating to the accretion of 
material from dense cores onto protostars remain poorly understood.  
As mass accretes onto protostars the gravitational energy is liberated and 
radiated away as accretion luminosity.  This luminosity, which depends on 
the mass accretion rate, current protostellar mass, and current protostellar 
radius, can be used to study the mass accretion process and distinguish 
between different accretion models.  

Observational studies of protostellar luminosities are hindered by the fact 
that protostars are deeply embedded in dense cores, with most of their 
emitted luminosities reprocessed to mid-infrared, far-infrared, and 
submillimeter wavelengths by the dust in the cores.  The first significant 
study of the protostellar luminosity distribution 
was presented in a series of papers by Kenyon et al.~(1990, 1994) 
and Kenyon \& Hartmann (1995).  They identified 23 protostars in the 
Taurus-Auriga molecular cloud and calculated bolometric luminosities by 
integrating the observed spectral energy distributions (SEDs) using 
\emph{Infrared Astronomical Satellite (IRAS)} $12-100$ \um\ photometry 
and longer-wavelength (sub)millimeter photometry from the ground, when 
available.  They found that the protostellar luminosity distribution extended 
from 0.09 -- 22 \lsun, with a mean and median of 2.3 \lsun\ and 0.7 \lsun, 
respectively, and a strong peak around 0.3 \lsun.

As first noted by Kenyon et al.~(1990), their observed protostellar 
luminosities are lower than expected from simple theoretical predictions.  
Their argument is as follows.  First, they assumed that all observed 
luminosity is accretion luminosity, 
\begin{equation}\label{eq_lacc}
L_{\rm acc} = f_{\rm acc} \frac{GM\dot{M}_{\rm acc}}{R} \qquad ,
\end{equation}
where $f_{acc}$ is an efficiency factor taken to be 1, $M$ and $R$ are 
the mass and radius of the protostar, and $\dot{M}_{acc}$ is the accretion rate 
onto the protostar.  By further assuming that the 
peak of the luminosity distribution is produced by low-mass stars with 
$M=0.1$ \msun\ and $R=1$ \rsun, they calculated an implied mass 
accretion rate of $\sim 10^{-7}$ \msun\ yr$^{-1}$.  If some fraction of the 
observed luminosity arises from the protostar itself (contraction, deuterium 
burning, etc.), the implied mass accretion rate is even lower.

In the simplest model, the collapse of a singular isothermal 
sphere initially at rest as first considered by Shu (1977) and later extended 
by Terebey, Shu, \& Cassen (1984) to include rotation (often called 
the ``standard model'' of star formation), collapse proceeds in an 
``inside-out'' fashion, beginning in the center of the core, moving outward at 
the sound speed, and giving rise to a constant mass accretion rate of 
$\dot{M}_{acc} \sim 2 \times 10^{-6}$ \msun\ yr$^{-1}$ for 10 K gas.  This is 
over ten times higher than inferred by Kenyon et al.~(1990), and will only 
scale upward as $\dot{M}_{acc} \propto T^{3/2}$ for higher gas temperatures.
Modifications to the standard 
model, including non-zero initial inward motions (Larson 1969; Penston 1969; 
Hunter 1977; Fatuzzo, Adams, \& Myers 2004), magnetic fields (Galli \& Shu 
1993a, 1993b; Li \& Shu 1997; Basu 1997), and isothermal spheres 
that are \emph{not} singular but feature flattened density profiles at small 
radii (Foster \& Chevalier 1993; Henriksen, \andre, \& Bontemps 1997) all 
tend to increase the accretion rate over that predicted by the standard model, 
making reconciliation between theory and the Kenyon et al.~observations 
difficult.  This has become known as the ``luminosity problem.''

Identification of protostars and determining their luminosities were both 
greatly improved by the launch 
of the \emph{Spitzer Space Telescope} (Werner et al.~2004) in 2003.  
Many sites of star formation have been observed at wavelengths ranging from 
3.6 to 160 \um\ through various \emph{Spitzer} surveys.  One such survey was 
the Legacy survey ``From Molecular Cores to Planet Forming Disks'' 
(hereafter c2d; Evans et al.~2003), which observed 7 large, nearby molecular 
clouds and $\sim$ 100 isolated dense cores and resulted in the discovery of 
very low luminosity objects (VeLLOs), protostars with internal 
luminosities\footnote{The internal 
luminosity, \lint, is the luminosity of the central source and excludes 
luminosity arising from external heating.} $\leq$ 0.1 \lsun\ embedded in dense 
cores (Young et al.~2004).  Dunham et al.~(2008) identified 15 VeLLOs in 
the c2d dataset, and detailed studies of several have confirmed their very 
low luminosities and status as embedded protostars (Dunham et al.~2006; Bourke 
et al.~2006; Lee et al.~2009; Dunham et al.~2010b; Kauffmann et al.~2011).

Both Enoch et al.~(2009) and Evans et al.~(2009) studied the c2d protostellar 
luminosity distribution by using the \emph{Spitzer} data to identify 
protostars, compiling complete SEDs including far-infrared and (sub)millimeter 
photometry from the literature, and integrating these SEDs to determine 
\lbol.  They found a total of 112 protostars in the seven c2d clouds.  Enoch 
et al.~calculated mean and median values similar to those found by Kenyon 
\& Hartmann (1995), but with their improved sample statistics they noted 
the presence of a larger fraction of sources at low luminosities ($\la 1.0$ 
\lsun).  Evans et al.~(2009) included a correction for foreground extinction 
and calculated revised mean and median values of 5.3 \lsun\ and 1.5 \lsun.  
In a separate study, Kryukova et al.~(2012) also 
derived the protostellar luminosity distribution for a number of star-forming 
clouds, including the c2d clouds.  They found an even larger excess of 
low-luminosity protostars than found by Enoch et al.~and Evans et al.  
Offner \& McKee (2011) argued that the higher observed luminosities found when 
extinction corrections are applied, combined with a more realistic value of 
the efficiency factor in Equation \ref{eq_lacc} of $f_{acc} \sim 0.5$ to 
take into account both the powering of jets and winds and the effects of 
unseen, episodic accretion bursts, can essentially resolve the luminosity 
problem, although explaining the large fraction of sources at very low 
luminosities remains a challenge.  

Several recent theoretical studies have explored possible resolutions to the 
luminosity problem, many of which were originally proposed by Kenyon et 
al.~(1990).  One possibility is that accretion is variable or episodic, 
with prolonged periods of low accretion punctuated by short bursts of rapid 
accretion.  Numerous origins for such a process have been proposed, including 
gravitational instabilities in protostellar disks (e.g., Vorobyov \& Basu 
2005, 2006, 2010; Machida et al.~2011; Cha \& Nayakshin 2011), a combination 
of gravitational and magneto-rotational instabilities in protostellar 
disks (e.g., Armitage et al.~2001; Zhu et al.~2009a, 2009b, 2010), 
quasi-periodic magnetically driven outflows in the envelope (Tassis \& 
Mouschovias 2005), decay and regrowth of MRI turbulence (Simon et al.~2011), 
close interaction in binary systems or in dense stellar clusters (Bonnell \& 
Bastien 1992; Pfalzner et al.~2008), and disk-planet interactions (Lodato \& 
Clarke 2004; Nayakshin \& Lodato 2011).  Indeed, Dunham \& Vorobyov (2012) 
showed that the \lbol\ distribution predicted by the Vorobyov \& Basu 
(2005, 2006, 2010) simulations, which feature highly variable accretion 
with episodic bursts, provides a reasonable match to the c2d observations 
presented by Evans et al.~(2009).  Alternatively, Offner \& McKee (2011) 
presented analytic derivations of the protostellar luminosity function for 
several different 
accretion scenarios and showed that accretion models that tend toward a 
constant accretion time rather than a constant accretion rate provide a good 
match to the Evans et al.~c2d observations.  As a third alternative, 
Dalba \& Stahler (2012) recently argued that external accretion onto 
collapsing cores from the surrounding background cloud will reduce accretion 
rates and luminosities.

\begin{deluxetable*}{lccll}
\tabletypesize{\scriptsize}
\tablewidth{0pt}
\tablecaption{\label{tab_clouds}Molecular Clouds Surveyed by the c2d and GB Surveys}
\tablehead{
\colhead{} & \colhead{} & \colhead{Distance} & \colhead{Distance} & \colhead{} \\
\colhead{Cloud} & \colhead{Survey} & \colhead{(pc)} & \colhead{Reference\tablenotemark{a}} & \colhead{Data Reference(s)\tablenotemark{b}}
}
\startdata
Aquila & GB       & 260 & Maury et al.~(2011) & Gutermuth et al.~(2008); Maury et al.~(2011)\\
Auriga/California & GB  & 450 & Lada et al.~(2009) & H.~Broekhoven-Fiene et al.~(2012, in preparation)\\
Cepheus           & GB  & 200--325\tablenotemark{c} & Kirk et al.~(2009) & Kirk et al.(2009)\\
Chamaeleon I      & GB  & 150 & Belloche et al.~(2011a) & \nodata\\
Chamaeleon II     & c2d & 178 & Whittet et al.~(1997) & Young et al.~(2005); Porras et al.~(2007); \alcala\ et al.~(2008)\\
Chamaeleon III    & GB  & 150 & Belloche et al.~(2011a) & \nodata\\
Corona Australis  & GB  & 130 & Neuh{\"a}user \& Forbrich (2008) & Peterson et al.~(2011)\\
IC5146            & GB  & 950 & Harvey et al.~(2008) & Harvey et al.~(2008)\\
Lupus I           & c2d & 150 & \comeron\ (2008) & Chapman et al.~(2007); \merin\ et al.~(2008)\\
Lupus III         & c2d & 200 & \comeron\ (2008) & Chapman et al.~(2007); \merin\ et al.~(2008)\\
Lupus IV          & c2d & 150 & \comeron\ (2008) & Chapman et al.~(2007); \merin\ et al.~(2008)\\
Lupus V           & GB  & 150 & \comeron\ (2008) & Spezzi et al.~(2011)\\
Lupus VI          & GB  & 150 & \comeron\ (2008) & Spezzi et al.~(2011)\\
Musca & GB & 160 & Knude \& Hog (1998) & T.~Huard et al.~(2012, in preparation)\\
Ophiuchus & c2d & 125 & de Geus et al.~(1989) & Padgett et al.~(2008)\\
Ophiuchus North & GB & 130 & Wilking et al.~(2008) & Hatchell et al.~(2012)\\
Perseus & c2d & 250 & Enoch et al.~(2006) & \jorgensen\ et al.~(2006); Rebull et al.~(2007)\\
Serpens & c2d & 429 & Dzib et al.~(2010, 2011) & Harvey et al.~(2006, 2007a, 2007b)
\enddata
\tablenotetext{a}{Reference for the distance quoted in this work.}
\tablenotetext{b}{References presenting the \emph{Spitzer} IRAC and MIPS observations.}
\tablenotetext{c}{Different regions within Cepheus are located at different distances; see Kirk et al.~(2009) for details.}
\end{deluxetable*}

With 112 protostars spread over more than three orders of magnitude in \lbol, 
the c2d sample of protostellar luminosities is still somewhat limited by small 
number statistics.  As a follow-up to c2d, the \emph{Spitzer} Gould Belt 
Legacy Survey (hereafter GB; L.~Allen et al.~2012, in preparation) observed 
most of the remaining clouds in the Gould Belt.  In this paper we extend 
the identification of protostars and calculations of \lbol\ from Evans 
et al.~(2009) to the combined c2d+GB dataset.  Our work is motivated by a 
desire for better underlying statistics in the observed protostellar luminosity 
distribution and improving the accuracy of the \lbol\ calculations by 
including additional data not yet available when the Evans et al.~study was 
conducted.  The organization of this paper is as follows:  We describe our 
method in \S \ref{sec_method}, including overviews of the c2d and GB surveys 
in \S \ref{sec_method_surveys}, the identification of protostars in \S 
\ref{sec_method_sample}, the compilation of full source SEDs in \S 
\ref{sec_method_seds}, and the calculation of \lbol\ in \S 
\ref{sec_method_calculate}.  \S \ref{sec_results} summarizes our basic 
results.  A discussion of these results is contained in \S 
\ref{sec_discussion}.  In particular, in \S \ref{sec_discussion_obs} we 
compare our results to the existing c2d (\S \ref{sec_discussion_c2d}) and 
Kryukova et al.~(2012) (\S \ref{sec_discussion_kryukova}) results, in \S 
\ref{sec_discussion_models} we discuss several recent theoretical 
investigations of protostellar luminosities, and in \S 
\ref{sec_discussion_firsmm} we evaluate the accuracy of our \lbol\ measurements 
for sources with observed SEDs that are not well sampled in the far-infrared 
and submillimeter, and the effects of this incomplete sampling on our 
overall results.  Finally, we outline important future work needed to further 
advance this topic in \S \ref{sec_future}, and summarize our findings in 
\S \ref{sec_summary}.

\section{Method}\label{sec_method}

\subsection{Overview of the Surveys}\label{sec_method_surveys}

The \emph{Spitzer} c2d survey (PI: N.~J.~Evans) conducted an imaging survey of 
seven large, nearby molecular clouds and about 100 isolated molecular cloud 
cores, and a spectroscopic survey of selected targets.  The science questions 
motivating this survey and a summary of the observation strategy are given by 
Evans et al.~(2003).  A summary of the results from the survey of the large 
molecular clouds is given by Evans et al.~(2009).  The 
\emph{Spitzer} GB survey (PI: L.~E.~Allen) was designed as a follow-up to the 
clouds portion of c2d and conducted an imaging survey of 11 nearby molecular 
clouds, completing most of the remaining clouds in the Gould Belt (L.~Allen et 
al.~2012, in preparation; see also Gutermuth et al.~2008; Harvey et al.~2008; 
Kirk et al.~2009; Peterson et al.~2011; Spezzi et al.~2011; Hatchell et 
al.~2012).  The two surveys obtained 3.6--8.0 \um\ images with the 
\emph{Spitzer} Infrared Array Camera (IRAC; Fazio et al.~2004) and 24--160 
\um\ images with the Multiband Imaging Photometer (MIPS: Rieke et al.~2004) of 
all 18 clouds.  A standard pipeline developed by c2d was used for data 
reduction, source extraction, and band-merging to produce final source 
catalogs for both surveys and has been described in detail elsewhere (Harvey 
et al.~2006; Evans et al.~2007).

Table \ref{tab_clouds} lists each cloud, the survey in 
which it was imaged (c2d or GB), the assumed distance to the cloud, the 
reference for the distance, and references of individual studies of each cloud 
where the observation strategy and basic results are presented.  These clouds 
were chosen to represent nearly all of the significant sites of star formation 
within the Gould Belt, with two major 
exceptions:  the Taurus and Orion molecular clouds.  These two clouds were 
each the focus of separate, dedicated \emph{Spitzer} Legacy surveys led by 
other groups, and folding in their results with the c2d+GB clouds will be the 
focus of a future paper.  The clouds listed in Table \ref{tab_clouds} span 
very large ranges of properties.  For example, the total cloud masses range 
from a few hundred \msun\ (e.g., Chamaeleon II; Evans et al.~2009) to 
$\sim 10^5$ \msun\ (Auriga/California Molecular Cloud; Lada et al.~2009), the 
star formation rates and star formation rate surface densities both span 
approximately two orders of magnitude (Evans et al.~2009; Heiderman et 
al.~2010), and the ratio of protostars to pre-main sequence stars 
(indicative of the amount of current star formation still on-going in the 
cloud) range from none (e.g., Lupus V and VI; Spezzi et al.~2011) to values 
in excess of 30\% (e.g., Auriga/California Molecular Cloud, Cepheus, IC5146, 
and Perseus; H.~Broekhoven-Fiene et al.~2012, in preparation; Kirk et al.~2009; 
Harvey et al.~2008; Evans et al.~2009).  We refer the reader to the 
individual cloud studies listed in Table \ref{tab_clouds} for more details 
and additional references.

We caution that the distances to the 18 clouds surveyed are not all 
well-known, and some cloud distances are still under significant debate.  One
such example is the debate over the distance(s) to the Serpens and 
Aquila regions.  Recent VLBA parallax measurements led to a 65\% increase in 
the distance to Serpens compared to that assumed by the c2d team (429 vs.~260 
pc; Strai{\v z}ys et al.~1996; Harvey et al.~2006; Dzib et al.~2010, 2011), 
and there remains debate whether or not Aquila is also located at this 
new, farther distance or even if all of Aquila is itself located at the same 
distance (e.g., Gutermuth et al.~2008; Maury et al.~2011).  We do not list 
formal distance uncertainties in Table \ref{tab_clouds} as such uncertainties 
are very poorly characterized in at least some clouds.  Instead we refer to the 
references listed in Table \ref{tab_clouds} for detailed discussions on the 
various methods used to derive distances and the uncertainties in these 
methods.  Future distance revisions will require future revisions to the 
results presented in this study.

\subsection{Sample Selection}\label{sec_method_sample}

Our method for selecting protostars from the c2d and GB observations closely 
follows the selection method used by Evans et al.~(2009) for the c2d clouds.  
We summarize the main points here and refer to Evans et al.~for more details.

The data reduction pipeline creates band-merged source catalogs incorporating 
2MASS and \emph{Spitzer} 1.25 -- 70 \um\ photometry for each cloud.  
Candidate young stellar objects (YSOs) are identified using a standard 
classification method developed for the \emph{Spitzer} c2d and GB projects.  
This method is described in detail in Harvey et al.~(2007b) and Evans et 
al.~(2007) and summarized in all of the publications presenting individual 
cloud studies listed in Table \ref{tab_clouds}.  Briefly, this method uses 
the \emph{Spitzer} SWIRE Legacy survey of the ELAIS N1 extragalactic field 
(Lonsdale et al.~2003), processed to simulate the sensitivity and extinction 
distribution of the clouds in the c2d and GB surveys, to determine the 
positions of galaxies in three different \emph{Spitzer} color-magnitude 
diagrams.  Each source extracted in the c2d and GB cloud catalogs with 
infrared colors indicative of the presence of dust (sources with colors that 
can not be explained by extincted background stars) is then assigned an 
unnormalized ``probability'' of being a galaxy or YSO based 
on its position in each color-magnitude diagram, its $K - $[4.5] color, 
whether it was found to be extended in either of the two shortest 
\emph{Spitzer} IRAC bands (3.6 and 4.5 \um), and its flux density at 24 
and 70 \um.  The color and magnitude boundaries, along with the final boundary 
between candidate YSO and candidate galaxy in unnormalized ``probability'', 
are set to provide a nearly 
complete elimination of SWIRE sources.  We refer the reader to Harvey 
et al.~(2007b) for further details on this classification method.  Similar 
classification methods have been presented by other \emph{Spitzer} studies 
of galactic star-forming regions (e.g., Gutermuth et al.~2009; Rebull et 
al.~2010; Kryukova et al.~2012).

In total, we identified 3239 candidate YSOs in the 18 c2d and GB catalogs.  
All sources were visually inspected to remove residual contaminants, 
including resolved galaxies misclassified as candidate YSOs and image 
artifacts identified as point-sources by the automated pipeline but lacking 
true point-source detections in one or more bands (see Evans et al.~2009 
for details).  Follow-up optical spectroscopy of targets in Serpens presented 
by Oliveira et al.~(2009) led to the identification and removal of 11 
background giants with infrared excesses.  We lack the data required to 
identify and remove such objects in the other clouds.  Oliveira et al.~(2009) 
found a contamination rate of 25\% in their Serpens study.  Serpens (and 
Aquila) are likely the worst cases due to their close proximity to the 
Galactic plane (spanning Galactic latitudes ranging from 2\degree\ to 
10\degree), although Romero et al.~(2012) recently suggested the contamination 
rate is at least as high in other clouds as well, and Hatchell et al.~(2012) 
found that 27\% of their sample of candidate YSOs in Ophiuchus North selected 
via the c2d criteria were likely to be background giants based on proper motion 
arguments.  However, 80\% of the contaminants identified by Oliveira et 
al.~and 75\% of the contaminants identified by Hatchell et al.~are classified 
as Class III YSOs, thus even if the overall contamination rate is as high as 
25\% -- 30\%, our inability to remove these contaminants will not 
significantly affect this study since it is only focused on the subset 
of YSOs that are considered to be protostars.  Finally, a few known YSOs 
missing from the list of candidate YSOs due to missing photometry at one or 
more \emph{Spitzer} wavelengths caused by saturation or nondetections from 
being too deeply embedded were added by hand.

The above process resulted in a final list of 2966 YSOs (since all 2966 sources 
passed visual inspection, we have followed the terminology used by Evans et 
al.~[2009] and dropped the word ``candidate'' at this point).  
This is nearly a factor of three 
increase over the 1024 YSOs identified in the c2d clouds alone by Evans et 
al.~(2009).  Many of these YSO populations have already been presented and 
discussed in detailed studies of individual clouds (see Table \ref{tab_clouds} 
for references) and in an analysis of the star formation rates 
and efficiencies of the c2d and GB clouds based on a preliminary version of the 
final YSO catalog (Heiderman et al.~2010).  A complete analysis of the 
full YSO population, implications for star formation rates and efficiencies in 
the Gould Belt, and the evolution and lifetimes of YSOs will be presented in a 
forthcoming paper (L.~Allen et al.~2012, in preparation).  Here we focus only 
on the observed luminosities of protostars.

The final sample of protostars is identified from the list of 2966 YSOs by 
examining the full SEDs compiled for each source (see below) and selecting 
only those sources associated with at least one (sub)millimeter detection 
at $\lambda \geq 350$ \um, resulting in a final sample of 230 protostars.  
This is identical to the procedure followed by 
Evans et al.~(2009) except they used a cutoff wavelength of 850 \um; we 
modified this to 350 \um\ because of the large increase in available data at 
this wavelength.  No intrinsic protostellar colors were assumed 
and no additional color criteria were imposed.  This decision is motivated by 
numerous recent studies that have used dust radiative transfer models to 
show that protostars observed through outflow cavities can resemble more 
evolved Class II or Class III sources in the infrared (e.g., Whitney et 
al.~2003; Robitaille et al.~2006; Crapsi et al.~2008; Dunham et al.~2010a).  
By selecting all sources associated with (sub)millimeter detections we 
recover such sources and identify all YSOs that are associated with dense 
cores, although future follow-up observations are required to remove true 
Class II or III sources simply seen in projection against a dense core.

By requiring a (sub)millimeter detection, our method requires the availability 
of (sub)millimeter surveys covering the full extents of the clouds surveyed 
by c2d and GB.  This is not always the case, as described in more detail 
in the next section below.  The effects of this limitation will be discussed 
in detail in \S \ref{sec_discussion_kryukova}, where we compare to a recent 
study that used very different methods for selecting protostars and did not 
require (sub)millimeter detections.

\subsection{Constructing Full SEDs and Correcting for Extinction}\label{sec_method_seds}

Similar to Evans et al.~(2009), we compiled as complete SEDs as possible for 
each of the 2966 YSOs.  In addition to the 2MASS and \emph{Spitzer} 1.25--70 
\um\ photometry provided by the source catalogs, we included the following:  
(1) optical photometry, where available from the literature, (2) 
\emph{Wide-field Infrared Survey Explorer (WISE)} 12 and 22 \um\ photometry 
from the all-sky catalog\footnote{Available at http://irsa.ipac.caltech.edu/cgi-bin/Gator/nph-scan?mission=irsa\&submit=Select\&projshort=WISE}, (3) selected 
other ground-based optical and infrared data as compiled by the authors of 
the detailed studies of individual clouds (see references in Table 
\ref{tab_clouds}), (4) \emph{Spitzer} 160 \um\ photometry for sources detected 
and not located in saturated or confused regions, calculated using aperture 
photometry and aperture corrections as given by the MIPS Instrument 
Handbook\footnote{Available at http://irsa.ipac.caltech.edu/data/SPITZER/docs/ 
mips/mipsinstrumenthandbook/}; (5) SHARC-II\footnote{The Submillimeter High 
Angular Resolution Camera II (SHARC-II) is a 350 \um\ bolometer array operated 
at the Caltech Submillimeter Observatory (Dowell et al.~2003).} 350 \um\ 
photometry, when available, from a targeted survey of protostellar sources 
(Wu et al.~2007; M.~M.~Dunham et al.~2012, in preparation); (6) 
SCUBA\footnote{The Submillimeter Common-User Bolometer Array (SCUBA) was a 450 
and 850 \um\ bolometer array operated at the James Clerk Maxwell Telescope.} 
450 and 850 \um\ photometry, when available, from the SCUBA Legacy Catalog 
(Di Francesco et al.~2008); and (7) other (sub)millimeter photometry from 
unbiased surveys of molecular clouds, where available.

For the last item above, other (sub)millimeter photometry from unbiased surveys 
of molecular clouds, we used photometry from the following surveys:  (1) A 
MAMBO2\footnote{The Max-Planck Millimeter Bolometer 2 (MAMBO2) was a 1.2 mm 
bolometer array operated at the IRAM 30-m telescope.} 1.2 mm survey of part of 
Aquila (Maury et al.~2011); (2) A LABOCA\footnote{The Large Apex Bolometer 
Camera (LABOCA) is an 870 \um\ bolometer array in operation at the Atacama 
Pathfinder Experiment telescope (Siringo et al.~2009).} 870 \um\ survey of 
Chamaeleon I (Belloche et al.~2011a); (3) A SIMBA\footnote{The SEST Imaging 
Bolometer Array was a 1.2 mm bolometer array in operation at the Swedish-ESO 
Submillimeter Telescope.} 1.2 mm survey of Chamaeleon II (Young et al.~2005); 
(4) A LABOCA 870 \um\ survey of Chamaeleon III (Belloche et al.~2011b); (5) A 
Bolocam \footnote{Bolocam is a 1.1 and 2.1 mm bolometer array operated at the 
Caltech Submillimeter Observatory (Glenn et al.~1998)} 1.1 mm survey of 
Ophiuchus (Young et al.~2006); (6) A Bolocam 1.1 mm survey of Perseus (Enoch 
et al.~2006); and (7) A Bolocam 1.1 mm survey of Serpens (Enoch et al.~2007).

Summarizing the above information, we have access to complete 
(sub)millimeter surveys for only 6 out of the 18 clouds (Chamaeleon I, 
Chamaeleon II, Chamaeleon III, Ophiuchus, Perseus, and Serpens), plus a 
partial survey ofAquila and piecemeal coverage of other clouds from the SCUBA 
Legacy Catalog (Di Francesco et al.~2008).  This incomplete (sub)millimeter 
coverage will affect both our luminosity calculations and ability to identify 
protostars, and these effects are discussed in detail in \S 
\ref{sec_discussion_firsmm} and \S \ref{sec_future_comprel}.

Finally, before using the SEDs to calculate bolometric luminosities, we 
correct the photometry for foreground extinction.  We wish to only correct 
for the foreground cloud extinction and not the local extinction from the 
dense core itself, as in the latter case the extincted emission is reprocessed 
to longer wavelengths and included in our observed SEDs.  Determining the true 
line-of-sight extinction to a protostar from the foreground cloud is not a 
trivial task.  Following Evans et al.~(2009), we assign extinction values 
to all 2966 YSOs (a sample which includes the 230 protostars identified in 
this work) as follows:
\begin{enumerate}
\item We adopt extinction values from the literature for Class II and III 
YSOs (classified via infrared spectral index; see Evans et al.~2009 for 
details) included in published optical studies.
\item We de-redden the remaining Class II and III YSOs to the intrinsic 
near-infrared colors of an assumed spectral type of K7, found to be fairly 
representative of the majority of Class II and III YSOs in the c2d clouds 
(Oliveira et al.~2009, 2010; see also Evans et al.~2009 for details).
\item We de-redden all of the Class I and Flat spectrum YSOs (again 
classified via infrared spectral index) in each cloud using the mean extinction 
toward all Class II YSOs in that cloud.
\end{enumerate}
The extinction values adopted for each of our 230 protostars following this 
procedure are listed in 
Table \ref{tabysoprops}.  Most of the protostars in a given cloud have the 
same adopted extinction value since most protostars are classified as Class I 
or flat spectrum via their infrared spectral index, although 
some have different values 
since no intrinsic protostellar colors were assumed by our selection criteria 
and thus some Class II YSOs are classified as protostars (see \S 
\ref{sec_method_sample} above).

Once the extinction values are assigned, we use these values combined with the 
Weingartner \& Draine (2001) extinction law for $R_V = 5.5$ to correct the 
photometry for extinction.  The choice of the $R_V = 5.5$ law rather than the 
$R_V = 3.1$ law is motivated by several studies showing that the former is more 
appropriate for the dense regions in which stars form (e.g., Chapman et 
al.~2009).  While we do caution that our approach is somewhat crude, it is the 
best that can currently be done and is significantly more reliable than 
ignoring the effects of extinction altogether.

\subsection{Calculation of Evolutionary Indicators}\label{sec_method_calculate}

Once we have constructed full SEDs as described above in \S 
\ref{sec_method_seds}, we use these SEDs to calculate the bolometric 
luminosities (\lbol) and bolometric temperatures (\tbol).  
\lbol\ is calculated by integrating over all detections,
\begin{equation}\label{eq_lbol}
\lbol = 4\pi d^2 \int_0^{\infty} S_{\nu}d\nu \qquad .
\end{equation}
The bolometric temperature is defined to be the temperature of a blackbody 
with the same flux-weighted mean frequency as the source (Myers \& Ladd 
1993).  Following Myers \& Ladd, \tbol\ is calculated as
\begin{equation}\label{eq_tbol}
\tbol = 1.25 \times 10^{-11} \, \frac{\int_0^{\infty} \nu 
S_{\nu} d\nu}{\int_0^{\infty} S_{\nu} d\nu} \quad \rm{K} \qquad .
\end{equation}
\tbol\ can be thought of as a protostellar equivalent of $T_{eff}$ 
for stars; \tbol\ 
starts at very low values ($\sim 10$ K) for cold, starless cores and 
eventually increases to $T_{eff}$ once the core and disk have fully 
dissipated.  The integrals defined in Equations \ref{eq_lbol} and \ref{eq_tbol} 
are calculated using the trapezoid rule to integrate over the finitely 
sampled SEDs.  To avoid model or fitting uncertainties and focus only on the 
observations themselves, we do not extrapolate beyond the shortest and longest 
frequences at which data are available and we do not interpolate over 
missing data.  Instead, we explore the effects of 
missing data on our \lbol\ calculations in \S \ref{sec_discussion_firsmm}.  
We calculate \lbol\ and \tbol\ twice, once with the original, observed 
photometry and once with the extinction-corrected photometry.  

\section{Results}\label{sec_results}

For each of the 230 protostars identified following the selection method 
described above, Table \ref{tabysoprops} lists 
a running index, the cloud in which the protostar is located, the 
\emph{Spitzer} source name (which also gives the coordinates), the assumed 
$A_V$ for extinction corrections, the infrared spectral index\footnote{The 
infrared spectral index, $\alpha$, is calculated over all 2MASS and 
\emph{Spitzer} detections from $2-24$ \um\ (Evans et al.~2007).} ($\alpha$), 
\tbol, and \lbol\ calculated from both the observed and extinction corrected 
photometry, and a flag indicating whether or not each protostar has any 
available data at 70 \um\ $< \lambda <$ 850 \um\ (see \S 
\ref{sec_discussion_firsmm}).  In Table \ref{tabysoprops} the extinction 
corrected values are denoted as \aprime, \tbolprime, and \lbolprime\ to 
differentiate them from the observed values.  Throughout the remainder of this 
paper we consider only the extinction corrected values and drop the primes for 
simplicity.  We do not give uncertainties for the \lbol\ derived in this work.  
Statistical uncertainties calculated by propagating through the uncertainties 
in the observed fluxes are on the order of 10\%, but the true uncertainties 
are dominated by incomplete sampling of the SEDs and are impossible to 
calculate for each source individually.  These uncertainties will be discussed 
further in \S \ref{sec_discussion_firsmm}.

\begin{figure}
\epsscale{1.0}
\plotone{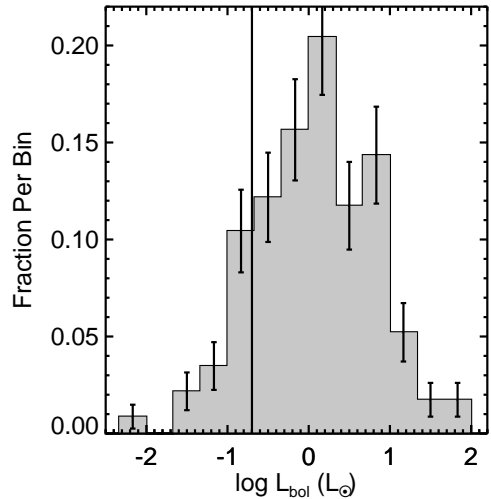}
\caption{\label{fig_lbol_hist}Histogram showing the distribution of extinction
corrected \lbol\ for all 230 protostars in log space.  The bins are 1/3 dex 
wide, and the error bars show the statistical ($\sqrt{N}$) uncertainties.  The 
solid vertical line shows the approximate completeness limit of 0.2 \lsun\ 
for the c2d+GB sample.}
\end{figure}

\begin{deluxetable*}{lr}
\tabletypesize{\scriptsize}
\tablewidth{0pt}
\tablecaption{\label{tab_stats}Luminosity Distribution Statistics}
\tablehead{
\colhead{Parameter} & \colhead{Value}
}
\startdata
Total Number & 230 \\
Mean & 4.3\tablenotemark{a} \lsun\ \\
Median & 1.3\tablenotemark{a} \lsun\ \\
Minimum & 0.01 \lsun\ \\
Maximum & 69 \lsun\ \\
Standard Deviation of log & 0.73 \\
Median / Mean & 0.3 \\
Maximum / Mean & 16.0 \\
Fraction $\leq 0.1$ \lsun\ & 0.07
\enddata
\tablenotetext{a}{As described in \S \ref{sec_discussion_firsmm}, once far-infrared and submillimeter photometry becomes available 
for the 43\% of the sample lacking any available data at 70 \um\ $< \lambda <$ 850 \um, the mean and median will likely increase 
to approximately 5.8 and 1.8 \lsun, respectively.  The effects of including such data on the overall distribution of \lbol, and thus 
on the other quantities listed in this Table, can only be investigated once such data are available.}
\end{deluxetable*}

Figure \ref{fig_lbol_hist} shows the distribution of the extinction corrected 
values of \lbol\ for all 230 protostars in log space.  With a minimum and 
maximum of 0.01 \lsun\ and 69 \lsun, respectively, this distribution extends 
over greater than three orders of magnitude.  The mean and median are 4.3 
\lsun\ and 1.3 \lsun, respectively.  These statistics are summarized in 
Table \ref{tab_stats}.  Also listed in Table \ref{tab_stats} are four 
dimensionless quantities calculated from the luminosity distribution: the 
standard deviation of log \lbol, the ratio of the median to mean \lbol, 
the ratio of the maximum to mean \lbol, and the fraction of protostars with 
\lbol\ $\leq 0.1$ \lsun.  These particular quantities are motivated by the 
recent theoretical study of protostellar luminosities by Offner \& McKee 
(2011), to which we compare our results below in \S 
\ref{sec_discussion_models}.

\begin{figure}
\epsscale{1.0}
\plotone{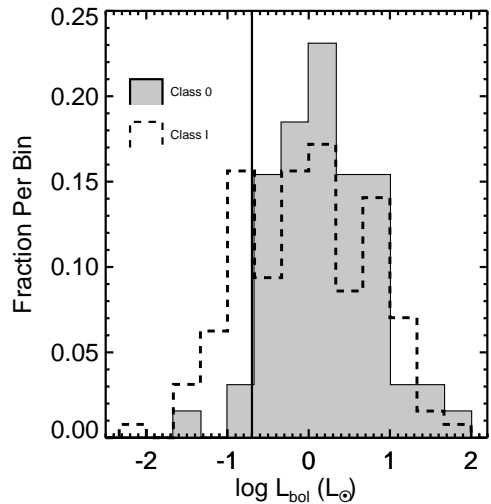}
\caption{\label{fig_lbol_hist_byclass}Histogram showing \lbol\ 
distributions in log space with 1/3 dex bins.  
The shaded histogram shows the distribution derived in this 
study for the 65 out of 230 objects in the combined c2d and GB samples 
classified as Class 0 protostars.  The dashed histogram shows the same thing, 
except for the 120 out of 230 objects classified as Class I protostars.  The 
classification is based on \tbol\ calculated according to Equation 
\ref{eq_tbol} using the extinction-corrected photometry and the Class 
boundaries defined by Chen et al.~(1995).  The solid vertical line shows the 
approximate completeness limit of 0.2 \lsun\ for the c2d+GB sample.}
\end{figure}

Figure \ref{fig_lbol_hist_byclass} shows the \lbol\ distributions separately 
for Class 0 and Class I sources.  We have used \tbol, calculated using 
Equation \ref{eq_tbol}, to classify our sources, since \tbol\ is one of 
several commonly used indicators of class and evolutionary status (e.g., 
Dunham et al.~2008; Enoch et al.~2009; Evans et al.~2009; Maury et al.~2011).  
Following Chen et al.~(1995), Class 0 sources are selected with the criterion 
that \tbol\ $< 70$ K and Class I sources are selected with the criterion that 
$70 \leq$ \tbol\ $\leq 650$ K.  Inspection of Figure 
\ref{fig_lbol_hist_byclass} reveals that the peak and extent of the \lbol\ 
distributions are similar for Class 0 and Class I sources.  The distributions 
have mean (median) values of 4.5 \lsun\ and 3.8 \lsun\ (1.4 \lsun\ and 1.0 
\lsun) for the Class 0 and I sources, respectively.  However, there is one 
significant difference in that there is an excess of low luminosity Class 
I sources compared to the Class 0 population.  For the Class I population, 
36\% have \lbol\ $< 0.5$ \lsun, whereas for the Class 0 population, only 
20\% have such luminosities.  A K-S test on the two distributions returns a 
value of only 0.04, demonstrating that the difference at low luminosities is 
statistically significant.  These results are similar 
to those obtained by Enoch et al.~(2009) for a smaller sample.  Very recently, 
several extremely low luminosity, Class 0 sources have been discovered in cores 
classified as starless based on \emph{Spitzer} observations (e.g., Chen et 
al.~2010; Enoch et al.~2010; Pineda et al.~2011; Schnee et al.~2012; 
Chen et al.~2012), emphasizing that at least some of this difference may be 
due to a bias against the lowest luminosity Class 0 sources in 
\emph{Spitzer}-selected samples.  This point is further emphasized by the fact 
that the excess of low-luminosity Class I sources occurs below our approximate 
completeness limit of 0.2 \lsun\ (see below), where any such comparisons are 
limited in utility.  The true similarity of the Class 0 and Class I \lbol\ 
distributions must be revisited once current and future surveys with 
\emph{Herschel} and ALMA detect and characterize the full population of 
extremely low luminosity protostars.

\begin{deluxetable}{lccc}
\tabletypesize{\scriptsize}
\tablewidth{0pt}
\tablecaption{\label{tab_vellos}\lbol\ and \lint\ for VeLLOs}
\tablehead{
\colhead{Source} & \colhead{\lbol} & \colhead{\lint} & \colhead{Reference\tablenotemark{a}}
}
\startdata
L1014-IRS & 0.34 & 0.09 & 1 \\
IRAM04191-IRS & 0.13 & 0.08 & 2 \\
L1521F-IRS & 0.13 & 0.05 & 3 \\
L328-IRS & 0.18 & 0.05 & 4 \\
L673-7-IRS & 0.18 & 0.04 & 5 \\
L1148-IRS & 0.13 & 0.08--0.13 & 6, 7
\enddata
\tablenotetext{a}{References: (1) Young et al.~(2004); (2) Dunham et al.~(2006); (3) Bourke et al.~(2006); 
(4) Lee et al.~(2009); (5) Dunham et al.~(2010b); (6) Kauffmann et al.~(2005); (7) Kauffmann et al.~(2011).}
\end{deluxetable}

We emphasize that the results presented here are the observed 
bolometric luminosities of protostars, which are not the same as the 
intrinsic protostellar luminosities.  
Departures from spherical symmetry break the correlation between observed 
and intrinsic bolometric luminosities, and external heating from the 
interstellar radiation field breaks the correlation between bolometric and 
internal luminosity.  
Regarding the latter, external heating can add up to several tenths of a solar 
luminosity depending on the local strength of the interstellar radiation field 
and the core mass available to be heated externally 
(e.g., Evans et al.~2001) and can dominate the observed \lbol\ for 
the lowest luminosity objects.  A few specific examples of this point can be 
found in recent, detailed studies of individual VeLLOs that use continuum 
radiative transfer models to separate internal and external heating and 
determine the intrinsic \lint.  The observed \lbol\ and model-derived \lint\ 
for six such sources are listed in Table \ref{tab_vellos}.  For at least 5 
and possibly all 6, the observed \lbol\ are above 0.1 \lsun\ while the 
model-derived \lint\ are below 0.1 \lsun, qualifying them as VeLLOs.  As a 
consequence, the fraction of protostars with \lbol\ $\leq$ 0.1 \lsun\ reported 
in Table \ref{tab_stats} (0.07) does not imply that 7\% of the sample are 
VeLLOs; many more VeLLOs with \lbol\ $> 0.1$ \lsun\ are likely present in the 
sample.

We have decided not to attempt to correct our luminosity distribution for 
source inclination or external heating, since any such corrections would be 
model-dependent (and in the case of external heating would require detailed 
modeling of all low-luminosity protostars, a project far beyond the scope of 
this paper).  What we present are simply the observed bolometric luminosities 
(after correcting for extinction).  Theoretical studies that attempt to explain 
the observed protostellar luminosity distribution must take these 
considerations into account.

Since the relationship between the fluxes in the various \emph{Spitzer} bands 
and \lbol\ depends not only on distance but also on the detailed spectral 
shape of each source, local strength of the external (interstellar) radiation 
field, and total core mass available to be heated externally, there is no one 
completeness limit for each cloud or for the full c2d+GB dataset.  In a 
detailed search for and study of low luminosity protostars in the c2d survey, 
Dunham et al.~(2008) found that the c2d data are sensitive to protostars with 
\lint\ $\geq 4 \times 10^{-3}$ $(d/140 \, \rm{pc})^2$ \lsun.  With cloud 
distances ranging from $125-950$ pc, the resulting luminosity sensitivites 
range from $0.003-0.18$ \lsun, or $0.003-0.04$ \lsun\ if IC5146 is omitted.  
However, this sensitivity is for \lint\ rather than \lbol; as discussed above, 
the two are not the same for low luminosity protostars, with \lbol\ equal to 
or greater than \lint\ depending on the details of the external heating.  In 
another study, Enoch et al.~(2009) estimated completeness limits of \lbol\ 
$\sim 0.01-0.05$ \lsun\ for protostars in the c2d clouds, although they 
emphasized that there was significant uncertainty in deriving such limits.  
Taking into account all of the above information, we conservatively estimate 
that our sample is only complete for \lbol\ $>$ 0.2 \lsun\ (the sensitivity 
limit for IC5146, the most distant cloud, using the Dunham et al.~[2008] 
relation and assuming no external heating), and mark this limit with a solid 
vertical line in all figures presenting histograms of \lbol.  The existence of 
protostars below these limits will be discussed in \S \ref{sec_future_lowlum}.

\section{Discussion}\label{sec_discussion}

In this section we discuss our results in comparison to other observational 
and theoretical studies of protostellar luminosities.  In \S 
\ref{sec_discussion_obs} we compare to two recent determinations of the 
observed protostellar luminosity distribution, and in \S 
\ref{sec_discussion_models} we discuss several recent theoretical 
investigations of protostellar luminosities.  Finally, in \S 
\ref{sec_discussion_firsmm} we discuss the effects of missing far-infrared 
and submillimeter photometry on our derived luminosities.

\subsection{Comparison to Other Observations}\label{sec_discussion_obs}

\subsubsection{Comparison to c2d Results}\label{sec_discussion_c2d}

Evans et al.~(2009) identified 112 protostars in the c2d survey and calculated 
their observed bolometric luminosities.  Our methods for identifying 
protostars, assembling complete SEDs, and calculating \lbol\ are very similar 
to theirs.  All of their protostars are included in our study, but we have 
expanded to the rest of the star-forming clouds observed by the GB survey and 
thus increased the number of protostars from 112 to 230.  We have also made 
three changes to the ancillary photometry included when assembling complete 
SEDs:  (1) we have included 12 and 22 \um\ photometry from the \emph{WISE} 
all-sky 
survey, (2) we have included additional SHARC-II 350 \um\ photometry from a 
targeted survey of nearby, low-mass star forming regions (Wu et al.~2007; 
M.~M.~Dunham et al.~2012, in preparation) that was not yet available when 
Evans et al.~(2009) completed their study; and (3) we have not included any 
\emph{IRAS} photometry.  The last change 
is motivated by the superiority of \emph{WISE} 12 and 22 \um\ and 
\emph{Spitzer} 70 \um\ data to \emph{IRAS} 12, 25, and 60 \um\ in essentially 
all cases, and the extreme confusion from both nearby sources and ambient 
cloud emission in the \emph{IRAS} 100 \um\ data.

\begin{figure}
\epsscale{1.0}
\plotone{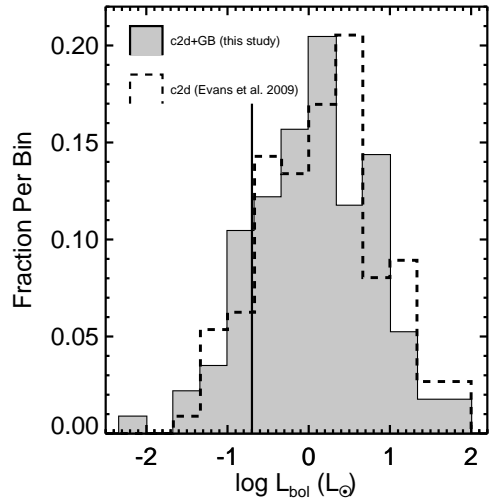}
\caption{\label{fig_lbol_hist_c2d}Histogram showing \lbol\ 
distributions in log space with 1/3 dex bins.  
The shaded histogram shows the distribution derived in this 
study for the 230 protostars in the combined c2d and GB samples (see Figure 
\ref{fig_lbol_hist} for error bars).  The dashed histogram shows the 
distribution for the 112 protostars in the c2d sample as derived by Evans 
et al.~(2009).  The solid vertical line shows the 
approximate completeness limit of 0.2 \lsun\ for the c2d+GB sample.}
\end{figure}

Figure \ref{fig_lbol_hist_c2d} compares the \lbol\ distributions from this 
work and from Evans et al.~(2009).  The new 
distribution obtained in this study has a similar shape and extent to the 
c2d-only distribution, except now with better statistics.  The medians are also 
quite similar, with values of 1.3 \lsun\ in this work and 1.5 \lsun\ in 
the c2d-only sample (Evans et al.~2009).  A K-S test on the two distributions 
returns a value of 0.33, indicating they are not significantly different.

\begin{figure}
\epsscale{1.0}
\plotone{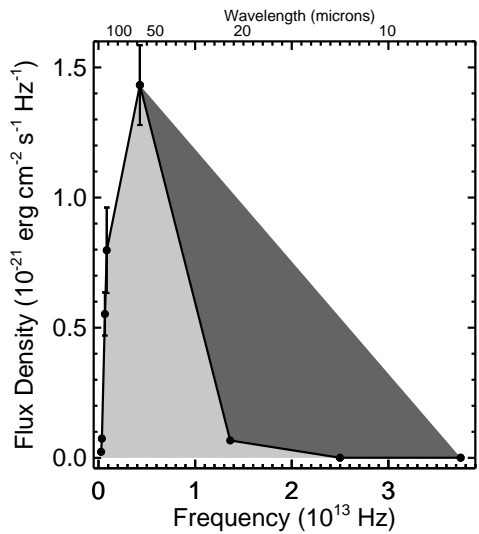}
\caption{\label{fig_sed_ngc1333_iras2a}Spectral energy distribution of 
NGC1333-IRAS2A, plotted as $S_{\nu}$ versus $\nu$ in linear space.  The light 
shaded area shows the result of integrating under the curve when the 
\emph{WISE} 12 and 22 \um\ photometry is included, whereas the dark shaded 
area shows the extra amount  amount added to the integral when no photometry 
is available between 8 and 70 \um, as was the case for Evans et al.~(2009).}
\end{figure}

Despite their general similarities, the two distributions do have different 
means:  4.3 \lsun\ in this study versus 5.3 \lsun\ in the c2d-only sample 
(Evans et al.~2009).  The mean is strongly influenced by the highest luminosity 
sources, several of which were overestimated by Evans et al.~(2009).  To 
understand the cause of this overestimate, we note that there are 14 sources in 
the Evans et al.~sample saturated at 24 \um\ with \emph{Spitzer} and thus 
lacking any photometry between 8 and 70 \um.  By including \emph{WISE} 12 and 
22 \um\ photometry, which was not available to Evans et al., our updated 
sample fills in this gap.  Figure \ref{fig_sed_ngc1333_iras2a} plots the SED 
for NGC1333-IRAS2A (source 144 in Table \ref{tabysoprops}), 
one of the 14 sources saturated in the \emph{Spitzer} 
24 \um\ observations.  The SED is plotted as $S_{\nu}$ versus $\nu$ in 
linear space rather than the more typical $\nu S_{\nu}$ versus $\lambda$ in log 
space since the former is the space in which the integral in Equation 
\ref{eq_lbol} is calculated.  The light shaded area shows the result of the 
integral when the \emph{WISE} 12 and 22 \um\ photometry is included, whereas 
the dark shaded area shows the extra amount added to the integral when 
no photometry is available between 8 and 70 \um.

As clearly demonstrated by Figure \ref{fig_sed_ngc1333_iras2a}, omitting 
photometry between 8 and 70 \um\ can lead to significant overestimates of 
\lbol.  In the specific case of NGC1333-IRAS2A, Evans et al.~(2009) measured 
\lbol\ $= 76$ \lsun\ whereas we measure \lbol\ $= 22$ \lsun\ with the 
\emph{WISE} 12 and 22 \um\ photometry included, a factor of 3.4 lower.  
Our measurement is consistent with previous measurements of \lbol\ for this 
source whereas the Evans et al.~value is not (e.g., \jorgensen\ et al.~2002).  
For the 14 sources saturated at 24 \um, Evans et al.~(2009) measured \lbol\ 
ranging from 2.6 to 76 \lsun, with a mean and median of 30 and 27 \lsun, 
respectively.  For those same 14 sources and with \emph{WISE} 12 and 22 \um\ 
photometry included, we measure \lbol\ ranging from 1.4 to 63 \lsun, with a 
mean and median of 20 and 13 \lsun, respectively.  Most of the decrease in 
the overall sample mean from 5.3 \lsun\ to 4.3 \lsun\ is a result of 
correcting this overestimate for several relatively high luminosity sources.

\subsubsection{Comparison to Kryukova Results}\label{sec_discussion_kryukova}

\begin{figure*}
\epsscale{1.0}
\plotone{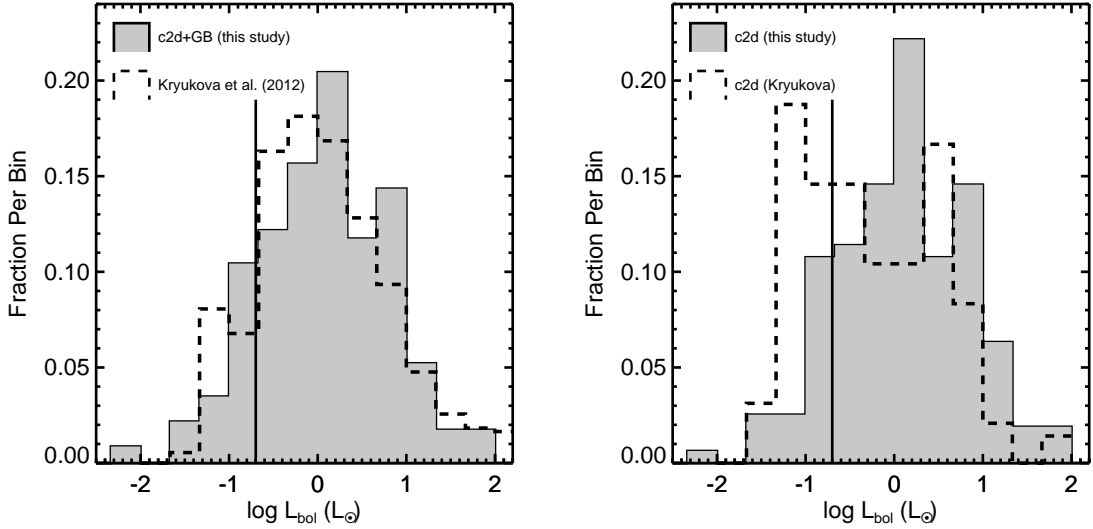}
\caption{\label{fig_lbol_hist_kryukova}Histogram showing \lbol\ distributions 
in log space with 1/3 dex bins.  In both panels, the solid vertical lines show 
the approximate completeness limits of 0.2 \lsun\ for the c2d+GB sample.
\emph{Left:}  The shaded histogram shows the distribution derived in this 
study for the 230 protostars in the combined c2d and GB samples (see Figure 
\ref{fig_lbol_hist} for error bars).  The dashed histogram shows the 
contamination-subtracted distribution for the 727 protostars identified by 
Kryukova et al.~(2012).  
\emph{Right:}  The shaded histogram shows the distribution from this work when 
only including the sources in the c2d clouds.  The dashed histogram shows 
the contamination-subtracted distribution from Kryukova et al.~(2012) when 
only including sources from the same c2d clouds.}
\end{figure*}

Recently, Kryukova et al.~(2012) presented observed protostellar luminosity 
distributions assembled from \emph{Spitzer} observations of 11 molecular 
clouds:  the 7 c2d clouds (Chamaeleon II, Lupus I, Lupus III, Lupus IV, 
Ophiuchus, Perseus, and Serpens), Taurus, and 3 massive star-forming 
clouds (Orion, Cep OB3, and Mon R2).  In total they identified 727 protostars 
in these clouds.  Figure \ref{fig_lbol_hist_kryukova} compares our results.

The left panel of Figure \ref{fig_lbol_hist_kryukova} compares the \lbol\ 
distributions from this work and from Kryukova et al.~(2012).  We use the 
contamination-subtracted \lbol\ distributions from Kryukova et al.~for this 
comparison. The two distributions are generally quite similar, but since 
Kryukova et al.~include three massive star-forming clouds in their sample, 
environmental effects may mask our ability to properly compare the two 
results.  Thus, the right panel of Figure \ref{fig_lbol_hist_kryukova} 
compares the \lbol\ distributions from this work and from Kryukova et al., 
where now both samples are restricted to the clouds common to both samples 
(the c2d clouds).

\begin{figure*}
\epsscale{1.0}
\plotone{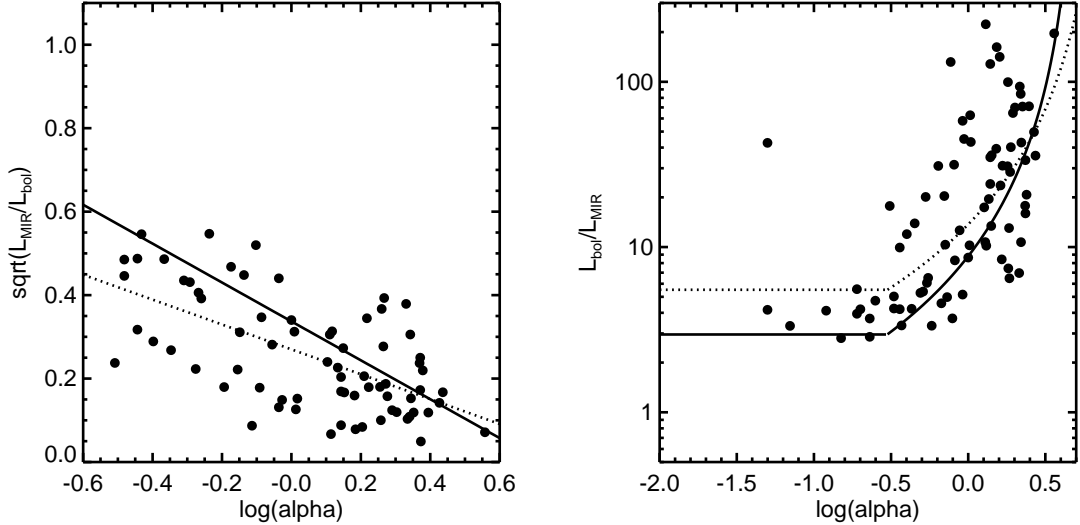}
\caption{\label{fig_alpha_lbol}Re-creation of Figure 5 from Kryukova et 
al.~(2012), showing $\sqrt{(\lmir/\lbol)}$ (\emph{left}) and \lbol/\lmir\ 
(\emph{right}) vs.~log($\alpha$) for the protostars common to both our 
sample and the Kryukova et al.~sample.  \lbol\ is from this work whereas 
\lmir\ (the mid-infrared luminosity) and $\alpha$ (the infrared spectral s
lope) are given by Kryukova et 
al.~(2012).  The solid lines show the best-fit relation from Kryukova et 
al., whereas the dotted line shows the modified relation derived using our 
new values of \lbol\ (see text for details).}
\end{figure*}

Inspection of the right panel of Figure \ref{fig_lbol_hist_kryukova} clearly 
shows that, for the same clouds, Kryukova et al.~(2012) find an observed 
distribution of protostellar luminosities that is generally shifted to 
lower luminosities compared to our results, with a much lower mean (2.3 
\lsun\ versus 4.2 \lsun\ in our sample) and much higher fraction of protostars 
with \lbol\ $\leq$ 0.1 \lsun\ (22\% versus 7\% in our sample).  
A K-S test on the two distributions returns a value of 0.01, verifying that 
the two distributions are statistically different.  
Kryukova et al.~do not compile complete SEDs to use in calculating 
bolometric luminosities.  Instead, for all sources they identify as protostars, 
they calculate \lmir, the mid-infrared luminosity from their 2MASS and 
\emph{Spitzer} $1.25-24$ \um\ data, and $\alpha$, the infrared spectral index 
calculated from 3.6 to 24 \um.  For sources with $\alpha \geq$ 0.3 common to 
both their sample of protostars and the Evans et al.~(2009) sample, they then 
derive the following empirical relationship:
\begin{equation}\label{eq_kryukova1}
\frac{\lmir}{\lbol} = (-0.466 \pm 0.014 \times \rm{log}(\alpha) + 0.337 \pm 0.053)^2 \qquad ,
\end{equation}
where \lbol\ is from Evans et al.~(2009).  They use this relation to calculate 
\lbol\ for all protostars with $\alpha \geq$ 0.3, and the value of this 
relation at $\alpha = 0.3$ to calculate \lbol\ for all protostars with 
$\alpha < 0.3$.  At least some of the discrepancy 
between our results and those of Kryukova et al.~may arise because we have 
made several changes to the SEDs used to calculate \lbol, as described above 
in \S \ref{sec_discussion_c2d}.  To examine this possibility, we re-derived 
the above empirical correlation using our new values of \lbol\ for the 
sources common to both our sample and the Kryukova et al.~sample, and obtained 
the following modification using a linear least-squares fit:
\begin{equation}\label{eq_kryukova2}
\frac{\lmir}{\lbol} = (-0.298 \pm 0.046 \times \rm{log}(\alpha) + 0.270 \pm 0.013)^2 \qquad .
\end{equation}
We illustrate the effects of this modification in Figure \ref{fig_alpha_lbol}, 
which re-creates Figure 5 from Kryukova et al.~(2012).

Using our modified relationship between \lbol, \lmir, and $\alpha$, the 
changes to the Kryukova et al.~(2012) \lbol\ values range from an increase 
by a factor of 1.9 to a decrease by a factor of two, depending on $\alpha$ for 
each source.  The mean change of all  
sources common to both samples is an increase by a factor of 1.5.  Such a 
change can explain much of the difference in means in the two samples (4.2 
\lsun\ in this study compared to 2.3 \lsun\ in Kyuokova et al.), but cannot 
fully explain the excess of low-luminosity sources.  Instead, the remainder 
of the discrepancy between our results lies in source selection.

\begin{figure}
\epsscale{1.0}
\plotone{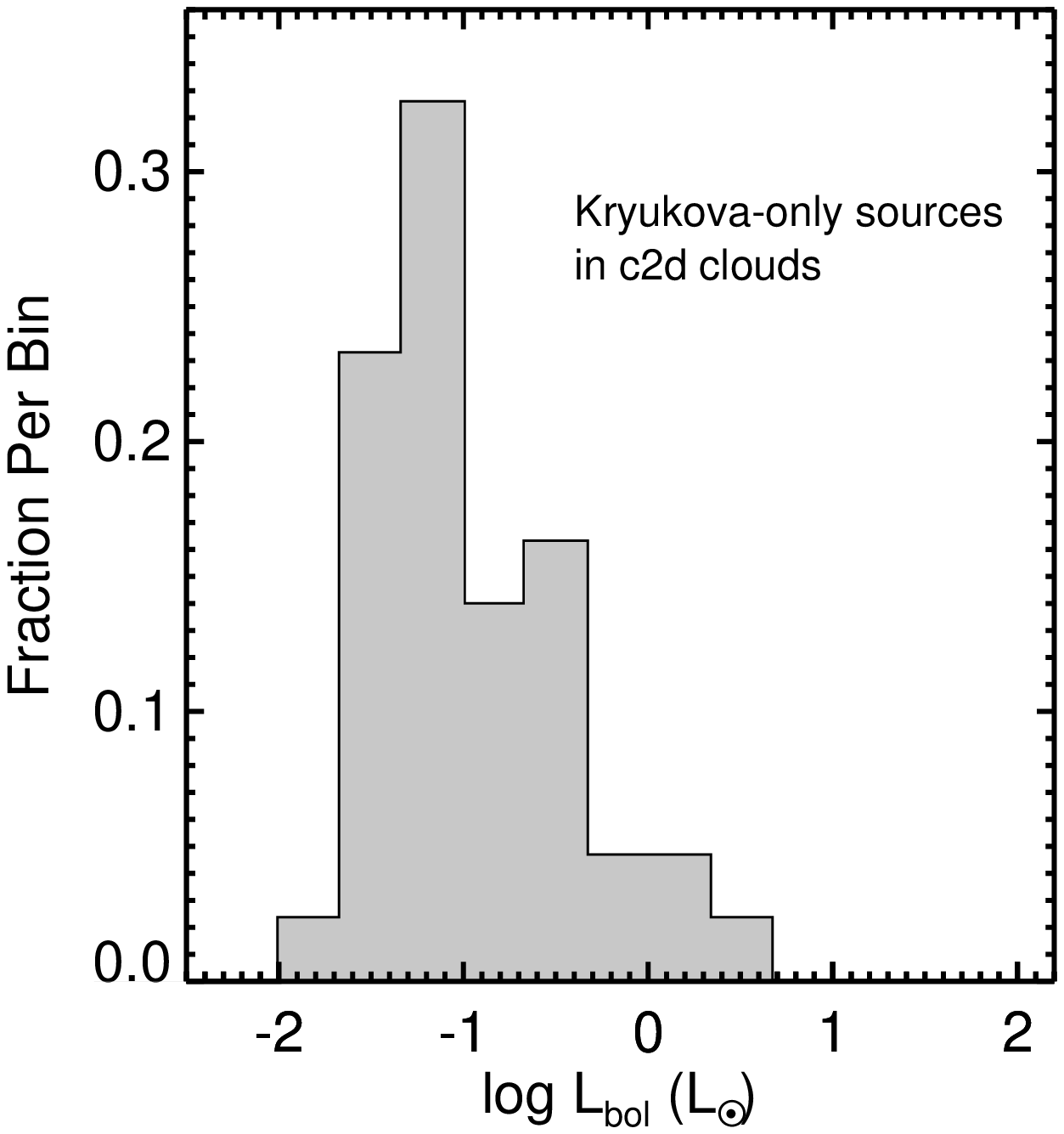}
\caption{\label{fig_lbol_kryukova_only}Histogram showing the \lbol\ 
distribution in log space for the 43 protostars in the c2d clouds identified 
by Kryukova et al.~(2012) but not by us or by Evans et al.~(2009).  The bins 
are 1/3 dex wide.}
\end{figure}

Kryukova et al.~identify 43 protostars in the c2d clouds not identified by us 
or by Evans et al.~(2009).  Figure \ref{fig_lbol_kryukova_only} shows the 
\lbol\ distribution for these 43 sources.  Most have \lbol\ $\leq 1.0$ \lsun, 
and while the effects of these ``extra'' low-luminosity sources are 
significantly mitigated by statistical contamination corrections included by 
Kryukova et al., their net effect is to cause an excess of low-luminosity 
sources compared to our results.  The main difference between our method of 
selecting protostars and that of Kryukova et al.~is our requirement of at 
least one detection at $\lambda \geq 350$ \um\ to ensure association with 
dense cores.  Since the 43 sources shown in Figure 
\ref{fig_lbol_kryukova_only} are not in our sample, they are not associated 
with (sub)millimeter detections and thus not associated with known dense 
cores.  By our definition of a protostar (see \S \ref{sec_intro}), these are 
not protostars.

However, it is possible that at least some of 
these sources are in fact protostars associated with relatively low-mass cores 
not detected by the (sub)millimeter surveys we used to compile complete SEDs.  
This is supported by the fact that many such surveys have relatively high 
completeness limits (for example, the 50\% completeness limits for the Bolocam 
1.1 mm surveys of Perseus, Serpens, and Ophiuchus are 0.8, 0.6, and 0.5 \msun, 
respectively; Enoch et al.~[2008]).  On the 
other hand, some sources may be contaminants masquerading in the 
sample.  Kryukova et al.~(2012) made a careful attempt to correct for such 
contamination in a statistical sense, but the resulting corrections are highly 
uncertain and may have been underestimated.  For example, they applied their 
protostar selection criteria to the catalog produced by the \emph{Spitzer} 
SWIRE Legacy survey of the ELAIS N1 extragalactic field (Lonsdale et al.~2003) 
to estimate the contamination from galaxies and remove the effects of this 
contamination from their luminosity distribution.  Such an estimate is a lower 
limit only because it does not take into account the fact that galaxies in 
their cloud source catalogs are observed through the extra extinction of the 
cloud itself, reddening all galaxies 
and thus increasing the number of galaxies with their assumed 
colors of protostars.  Furthermore, Heiderman et al.~(2010) 
recently showed that many sources selected as Class I YSOs by their infrared 
colors lacked the presence of warm, dense gas and are thus not protostars (and 
may not even be YSOs at all; see Heiderman et al.~for details).  Since 
Kryukova et al.~used similar color-based selection criteria, it is plausible 
that not all of their sources are actually protostars.  Our 
(sub)millimeter detection requirement should remove such fake sources.

Ultimately, we conclude that there are limitations to both our method and 
that used by Kryukova et al.  We are limited by the availability and 
sensitivity of (sub)millimeter surveys whereas they are limited by uncertain 
statistical corrections for contamination.  We prefer our method because it 
ensures a reliable sample, but the Kryukova et al.~(2012) results emphasize 
that this reliability may come at the expense of completeness.  Some 
of the ``extra'' 43 sources in the c2d clouds they identify but we do not may 
in fact be real protostars.  Quantifying how many is simply not possible until 
future surveys in the far-infrared and submillimeter become available, as 
discussed further in \S \ref{sec_future_comprel}.  

\subsection{Theoretical Investigations of Protostellar Luminosities}\label{sec_discussion_models}

With the new observed protostellar luminosity distributions derived from large 
\emph{Spitzer} surveys (Evans et al.~2009; Kryukova et al.~2012; this work), 
several recent studies comparing the predictions of theoretical accretion 
processes to the observations have been published.  In one such study, Dunham 
\& Vorobyov (2012) coupled two-dimensional radiative transfer calculations with 
the numerical hydrodynamical simulations of Vorobyov \& Basu (2005, 2006, 
2010).  These simulations predict accretion rates that both 
generally decline with time and feature short-term variability and episodic 
bursts caused by disk gravitational instability and fragmentation.  Dunham \& 
Vorobyov (2012) used the core, disk, and protostellar masses, radii, and mass 
accretion rates predicted by the simulations as inputs to their radiative 
transfer calculations.  They included the effects of external heating 
in their radiative transfer models, and calculated model SEDs at all 
inclinations from the beginning of collapse until the end of the embedded 
phase.  They used these SEDs to calculate \lbol\ in the same manner as 
observers (Equation \ref{eq_lbol}) at all inclinations and all timesteps.  
Finally, they assembled a theoretical prediction of the observed luminosity 
distribution by calculating the fraction of total time the models spend at 
each \lbol, weighted by inclination and initial core mass.

\begin{figure}
\epsscale{1.0}
\plotone{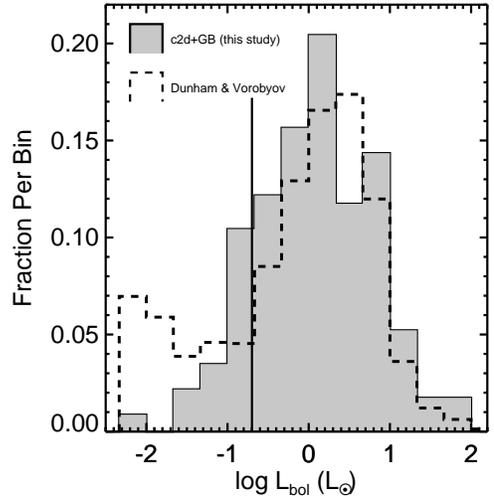}
\caption{\label{fig_lbol_hist_dunham}Histogram showing \lbol\ 
distributions in log space with 1/3 dex bins.  
The shaded histogram shows the distribution derived in this 
study for the 230 protostars in the combined c2d and GB samples (see Figure 
\ref{fig_lbol_hist} for error bars).  The dashed histogram shows the 
model distribution derived by Dunham \& Vorobyov (2012) based on a set of 
hydrodynamical simulations predicting accretion rates that both 
generally decline with time and feature short-term variability and episodic 
bursts caused by disk gravitational instability and fragmentation.  The solid 
vertical line shows the approximate completeness limit of 0.2 \lsun\ for the 
c2d+GB sample.}
\end{figure}

Dunham \& Vorobyov compared their model luminosity distribution to the 
c2d observations presented by Evans et al.~(2009) and showed that the models 
generally match the shape and spread of the observed luminosity distribution, 
indicating that the underlying variable accretion process predicted by the 
Vorobyov \& Basu (2005, 2006, 2010) simulations offers a possible resolution 
of the luminosity problem.  Figure \ref{fig_lbol_hist_dunham} compares our 
new observations with the Dunham \& Vorobyov (2012) models.  As our results 
are generally similar to Evans et al.~(2009) in terms of the shape and extent 
of the observed luminosity distribution (see \S \ref{sec_discussion_c2d}), 
except with improved statistics, the basic conclusions of Dunham \& Vorobyov 
are unchanged.  While the agreement is not perfect, the models provide a 
reasonable match to the observed luminosity distribution, with a K-S test on 
the two distributions returning a value of 0.41.  The only significant 
discrepancy between our observations and the Dunham \& Vorobyov (2012) models 
is at \lbol\ $\la 0.2$ \lsun.  As discussed in detail by Dunham \& 
Vorobyov (2012), this \lbol\ regime is at or below the completeness limit of 
the c2d and GB surveys, rendering a proper comparison of observed and model 
\lbol\ impossible at such luminosities (see also \S \ref{sec_future_lowlum}).  
The variable accretion process predicted by the Vorobyov 
\& Basu (2005, 2006, 2010) simulations and considered by Dunham \& Vorobyov 
(2012) remains a valid solution to the luminosity problem.

Offner \& McKee (2011) have also presented a recent theoretical study of 
protostellar luminosities.  They derived analytic luminosity functions for 
several different accretion scenarios and compared their results to the 
c2d observations presented by Evans et al.~(2009).  In addition to  
comparing to the mean and median of the observed distribution, they calculated 
four dimensionless quantities to characterize their luminosity functions and 
compared to the same quantities calculated from the observations.  Their study 
was our motivation for calculating and tabulating the quantities listed in 
the last four rows of Table \ref{tab_stats}, although we emphasize that the 
last quantity presented in Table \ref{tab_stats}, the fraction of protostars 
with \lbol\ $\leq 0.1$ \lsun, is not the same as the fraction of VeLLOs from 
Offner \& McKee (2011).  Their quantity was calculated as the ratio of sources 
with \lbol\ $\leq 0.14$ \lsun\ to those with \lbol\ $\leq 1.4$ \lsun\ and was 
chosen to provide a direct comparison to the observations presented by 
Dunham et al.~(2008).

Based on comparing their analytic luminosity functions to the c2d observations, 
Offner \& McKee (2011) concluded that accretion scenarios that tend toward 
a constant accretion time rather than a constant accretion rate are better 
able to match the observed protostellar luminosity distribution.  Given the 
similarity of our observed luminosity distribution to that of the c2d sample, 
the new observations presented here do little to change 
the findings of Offner \& McKee.  Their conclusion is in general agreement with 
a series of investigations by Myers (2010, 2011, 2012), who derived 
analytic luminosity distributions based on simple models of protostellar 
evolution assuming constant protostellar birth rates, accretion from both the 
dense core and from the surrounding ambient medium (``core-clump'' accretion, 
to use their terminology), and accretion durations set by the assumption of 
an equally likely stopping time.  They showed that such models, which predict 
accretion rates that increase with protostellar mass and thus tend toward a 
constant accretion time rather than constant accretion rate, exhibit good 
agreement with observed protostellar luminosities.

While the Dunham \& Vorobyov (2012), Offner \& McKee (2011), and Myers 
(2010, 2011, 2012) models have all succeeded in matching the observed 
distribution of protostellar luminosities, they do so with models featuring 
very different accretion properties.  Like the collapse of a singular 
isothermal sphere, which does not match observations, the Dunham \& 
Vorobyov (2012) models feature \emph{time-averaged} accretion rates (averaged 
over the full duration of the embedded phase) that do not vary with 
the final mass of the protostar.  In other words, higher mass protostars 
require more time to form than low-mass protostars.  Their solution to the 
luminosity problem is to invoke variaiblity and episodic accretion bursts as 
predicted by the Vorobyov \& Basu (2005, 2006, 2010) simulations, which 
reduce the accretion rates for most times and increase them during short-lived 
accretion bursts.  On the other hand, Offner \& McKee (2011) and Myers 
(2010, 2011, 2012) solve the luminosity problem by adopting models with 
time-averaged accretion rates that increase with the final mass of a 
protostar, so that all protostars form in about the same amount of time 
regardless of their final mass.  These studies emphasize that there is more 
than one possible resolution to the luminosity problem, and future theoretical 
work is needed to better decipher the implications of protostellar 
luminosities for the underlying mass accretion process and to distinguish 
between these different accretion scenarios.  In this work we have assembled a 
larger dataset to which such work should compare.

One weakness of both the Offner \& McKee (2011) and Myers (2010, 2011, 2012) 
models is that they do not predict \emph{observed} protostellar luminosities.  
The observed luminosity of a protostar includes accretion luminosity, 
photosphere luminosity, and external luminosity from heating by the 
interstellar radiation field, and is dependent on the physical structure and 
inclination of the source.  Offner \& McKee (2011) only include the two 
internal luminosity components (accretion and photosphere), and Myers 
(2010, 2011, 2012) include only accretion luminosity.  Furthermore, neither 
set of models takes into account the effects of source structure and 
inclination.  The evolutionary radiative transfer models presented by 
Dunham \& Vorobyov (2012) do take all these effects into account and calculate 
observed luminosities before comparing to observations, but they can also be 
criticized for assuming a very simple physical structure, not allowing for any 
variation in this structure, adopting a fixed interstellar radiation field 
with no variation in its strength, attenuation, or spectral shape, and 
weighting by only one of several formulations of the stellar initial mass 
function.  Future theoretical work must build on the foundations laid by these 
recent studies to properly compare theoretical predictions to observed 
protostellar luminosities.

\subsection{Effects of Including FIR/SMM Photometry}\label{sec_discussion_firsmm}

The SEDs of embedded sources typically peak around $100-300$ \um\ (e.g., 
\andre\ et al.~1999; Enoch et al.~2009; Dunham et al.~2008); accurate 
sampling in the far-infrared and submillimeter is thus necessary to ensure 
accurate measurements of \lbol.  Missing photometry near the peaks of the SEDs 
will result in underestimates of \lbol.  As described above in \S 
\ref{sec_method_seds}, we included \emph{Spitzer} 160 \um\ photometry for 
sources detected and not located in saturated or confused regions, SHARC-II 
350 \um\ photometry when available from a targeted survey of protostellar 
sources (Wu et al.~2007; M.~M.~Dunham et al.~2012, in preparation), and SCUBA 
450 \um\ photometry when available in the SCUBA Legacy Catalog (Di Francesco 
et al.~2008).  Out of the 230 total protostars identified in this work, 130 
(57\%) include at least one photometry point at 160, 350, or 450 \um.  
The other 100 (43\%) lack any available photometry between 
70 \um\ and at least 850 \um, and sometimes between 70 \um\ and 1.1 mm.  
The last column of Table \ref{tabysoprops} indicates whether each protostar 
has any such photometry available.

\begin{figure*}
\epsscale{1.0}
\plotone{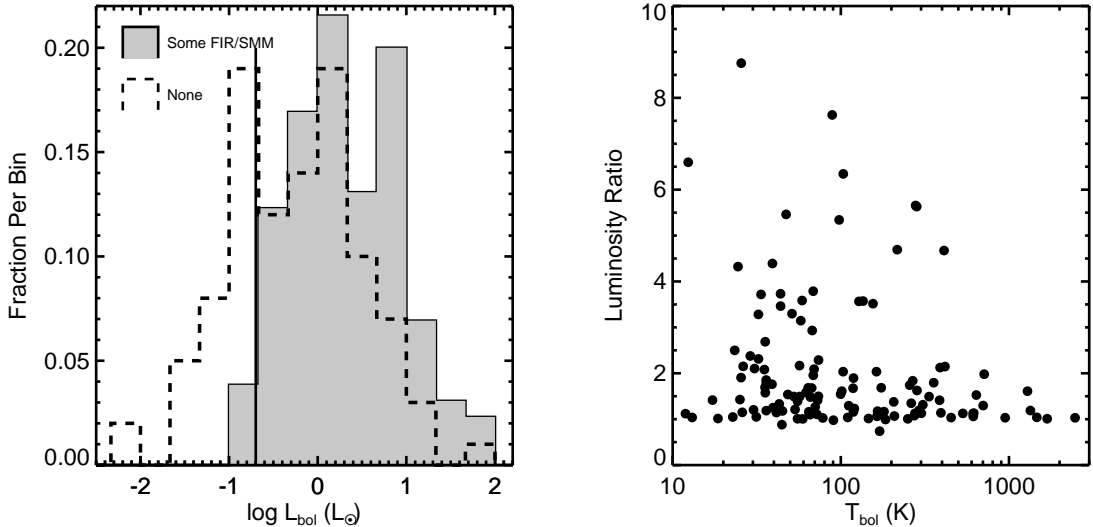}
\caption{\label{fig_fir_smm}\emph{Left:}  
\lbol\ histograms for the 230 protostars from this 
study in log space with 1/3 dex bins.  
The solid, shaded histogram shows the \lbol\ distribution for the 
130 protostars with at least one detection at 160, 350, or 450 \um, whereas 
the dashed histogram shows the \lbol\ distribution for the 100 protostars 
with no available photometry between 70 and 850 \um.  The solid 
vertical line shows the approximate completeness limit of 0.2 \lsun\ for the 
c2d+GB sample.  
\emph{Right:}  The ratio of \lbol\ calculated with all available photometry 
to \lbol\ calculated with all detections between 70 and 850 \um\ removed 
for the 130 protostars with at least one detection in this wavelength range, 
plotted versus \tbol.}
\end{figure*}

How much have we underestimated \lbol\ for the 43\% of the sample lacking any 
far-infrared and submillimeter photometry, and what effect does this have on 
the derived luminosity distribution?  To address these questions, the left 
panel of Figure \ref{fig_fir_smm} plots the \lbol\ distribution separately 
for the sources with and without at least one observed photometry point
 at 160, 350, or 450 
\um.  As expected, the \lbol\ distribution for sources without any far-infrared 
or submillimeter data is shifted to lower luminosities.  To quantify 
the amount by which \lbol\ is underestimated for these sources, we took the 
130 protostars with at least one such observed photometry point, 
removed all data between 
70 and 850 \um, and recalculated \lbol.  The results are shown in the right 
panel of Figure \ref{fig_fir_smm}, which plots the ratio of \lbol\ calculated 
with all available photometry to \lbol\ calculated with all data between 
70 and 850 \um\ removed versus \tbol.  The mean and median of this ratio are 
2.6 and 1.5, with a few sources showing underestimates in \lbol\ up to 
factors of $8-10$.  There may be some evidence for a trend of larger \lbol\ 
underestimates for more deeply embedded sources (those with lower \tbol).  
Such a trend is not surprising since a greater fraction of the total luminosity 
is emitted in the far-infrared and submillimeter for more deeply embedded 
sources, but, if present, the trend is not very strong.  Even sources 
with a \tbol\ of several hundred K can have \lbol\ underestimated by 
a factor of 2 or more.

\begin{figure}
\epsscale{1.0}
\plotone{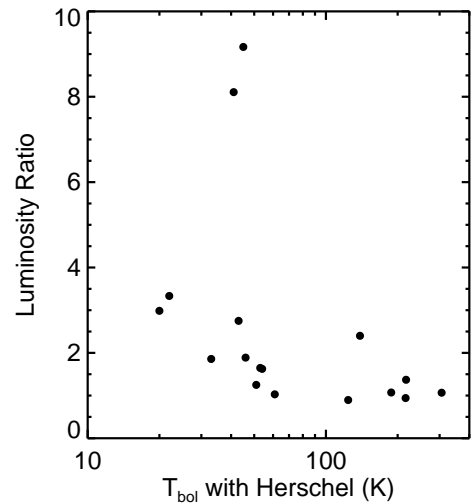}
\caption{\label{fig_herschel_aquila}The ratio of \lbol\ with 
\emph{Herschel} photometry to \lbol\ without \emph{Herschel} 
photometry versus \tbol\ calculated with \emph{Herschel} photometry for the 
17 protostars in Aquila.  
\lbol\ and \tbol\ with \emph{Herschel} photometry are taken from Maury et 
al.~(2011) and \lbol\ without \emph{Herschel} photometry is calculated in this 
work with the observed photometry (see text for details).}
\end{figure}

Another way of examining the effects of missing far-infrared and submillimeter 
data is to use \emph{Herschel} $100-500$ \um\ photometry.  The \emph{Herschel} 
Gould Belt survey is currently surveying all of the clouds included here and 
will eventually provide such photometry (\andre\ et al.~2010).  
Maury et al.~(2011) recently used 
data from this survey to study the protostellar population of part of 
Aquila.  The ideal comparison would be to include \emph{Herschel} photometry 
in our SEDs and recalculate \lbol\ for each source, but we are unable to do 
this since Maury et al.~did not publish the \emph{Herschel} fluxes.  They did, 
however, publish their own measurement of \lbol\ for each source including 
the \emph{Herschel} data.  We have identified 17 protostars in Aquila that 
overlap with their sample; in all 17 cases we lacked any far-infrared or 
submillimeter photometry.  Figure \ref{fig_herschel_aquila} plots the ratio 
of \lbol\ from Maury et al.~(2011) with the \emph{Herschel} data included 
to \lbol\ from this work without the \emph{Herschel} data versus \tbol\ 
from Maury et al.~(2011) with the \emph{Herschel} data included.  
The results are strikingly similar to our above analysis: including the 
\emph{Herschel} far-infrared and submillimeter photometry increases \lbol\ 
by a mean and median of 2.6 and 1.6, respectively, and by up to factors of 
$8-10$ in some cases.  Furthermore, there is again some evidence of a trend 
with \tbol, but the sample size is too small for robust statistics.  
This analysis should be repeated for a larger sample and with \lbol\ 
calculated in a consistent manner with and without the \emph{Herschel} 
photometry, but the agreement with the above analysis is encouraging.

Based on two separate investigations, we conclude that lacking any far-infrared 
or submillimeter photometry will lead to underestimates in \lbol\ by a 
factor of about 2.5 on average, but up to factors of $8-10$ in extreme cases.  
These findings are consistent with earlier results by Dunham et al.~(2008) and 
Enoch et al.~(2009), who also investigated this topic.  
Since nearly half (43\%) of our sample currently lacks any such photometry, 
including \emph{Herschel} photometry when it becomes available will affect our 
derived luminosity distribution.  To determine how significant this effect 
will be, we increased the calculated \lbol\ for all 100 sources lacking 
any far-infrared or submillimeter photometry by factors of 2.5 and found 
that the mean and median of the full sample of 230 sources 
increase to 5.8 and 1.8 \lsun, respectively.  Compared 
to the values listed in Table \ref{tab_stats}, both quantities increase by 
35\% -- 40\%.  Testing this conclusion and determining the full effects on 
the shape and extent of the luminosity distribution can only be done once the 
\emph{Herschel} Gould Belt survey is complete and the photometry is available.

\section{Future Work}\label{sec_future}

\subsection{Completeness and Reliability of the Sample}\label{sec_future_comprel}

In order to be classified as a protostar, we require at least one detection 
at $\lambda \geq 350$ \um\ to signify the presence of a dense core.  However, 
as described in \S \ref{sec_method_seds}, we only have access to complete 
(sub)millimeter surveys of 6 out of the 18 clouds (Chamaeleon I, Chamaeleon 
II, Chamaeleon III, Ophiuchus, Perseus, and Serpens), plus a partial survey of
Aquila (Maury et al.~2011) and piecemeal coverage of other clouds from the 
SCUBA Legacy Catalog (Di Francesco et al.~2008).  We are almost certainly 
missing protostars from the clouds where we have incomplete (sub)millimeter 
coverage.  Furthermore, where we do have coverage, the surveys are typically 
incomplete to low-mass cores with $M \la 0.5$ \msun\ (e.g., Enoch et 
al.~2008).  We are also likely missing some protostars in low-mass cores.

Kryukova et al.~(2012) have also studied the protostellar luminosity 
distribution.  They do not require a (sub)millimeter detection to be included 
in their sample.  In a detailed comparison between our results and their 
results (see \S \ref{sec_discussion_kryukova}) we showed that they identify 
approximately 40\% more protostars than we do in the same clouds and that these 
``extra'' protostars mostly have \lbol\ $\leq 1.0$ \lsun.  Some may be true 
protostars located in low-mass cores below the sensitivities of the 
(sub)millimeter surveys available to us, but some may also be contaminants 
masquerading as protostars in their sample.

Based on the presently available data, we are unable to evaluate how 
significantly our sample lacks completeness and how significantly the Kryukova 
et al.~(2012) sample lacks reliability.  Both should be fully resolved once 
the JCMT SCUBA-2 Legacy (Ward-Thompson et al.~2007) and \emph{Herschel} 
(\andre\ et al.~2010) Gould Belt Surveys are complete, since the two surveys 
together will fully characterize the population of dense cores in all of the 
Gould Belt clouds with sensitivities below 0.1 \msun.  Future work must 
revisit the topic of defining the sample of protostars once the results from 
these surveys are available.

\subsection{Accuracy of \lbol\ Calculation}\label{sec_future_lbol}

As discussed in detail in \S \ref{sec_discussion_firsmm}, nearly half (43\%) 
of our sample lacks any detections between 70 \um\ and at least 850 \um, and 
sometimes between 70 \um\ and 1.1 mm.  We have shown that such sources will 
have their \lbol\ underestimated by a factor of about 2.5, on average, and 
up to factors of $8-10$ in extreme cases.  We have estimated that correcting 
these underestimates will increase both the mean and median protostellar 
luminosity by about 35\% -- 40\%.  Verifying this estimate and examining the 
full effects of these underestimates on the shape and extent of the luminosity 
distribution must be revisited once the \emph{Herschel} Gould Belt survey is 
complete and provides $100-500$ \um\ photometry for our full sample.

\subsection{Extending to Lower Luminosities}\label{sec_future_lowlum}

As discussed in \S \ref{sec_results}, our protostellar sample is generally 
only complete for \lbol\ $> 0.2$ \lsun, with the exact completeness limit 
different for each source depending on its distance, detailed spectral shape, 
and amount of external heating.  Does there exist a 
population of protostars with luminosities below the completeness of the c2d 
and Gould Belt surveys?

Recent evidence suggests that the answer is yes.  First, in a detailed study of 
the population of low luminosity protostars in the c2d clouds, Dunham et 
al.~(2008) noted that protostars were found all the way down to the 
sensitivity limit and suggested that the lower limit to protostellar 
luminosities had not yet been found.  Second, very recent work has identified 
a population of cores originally believed to be starless and undetected in 
c2d and other \emph{Spitzer} observations of similar depth but found to be 
driving molecular outflows through sensitive (sub)millimeter interferometer 
observations.  To date, seven such objects have been identified, and all but 
one are located in the Perseus Molecular Cloud (Chen et al.~2010; Enoch et 
al.~2010; Dunham et al.~2011; Pineda et al.~2011; Schnee et al.~2012; Chen 
et al.~2012; Pezzuto et al.~2012).

Both the true number and evolutionary status of these objects remains 
unknown.  Most have been discovered serendipitously through (sub)millimeter 
interferometer detections of outflows in observations targeting cores believed 
to be starless.  A full survey for such objects has not been possible to date 
due to the prohibitively large time requests that would be required, although 
such surveys should be possible in the very near future with ALMA.  None of 
the sources were detected in the infrared with c2d or other \emph{Spitzer} 
surveys of similar depth, implying upper limits of 0.01 -- 0.1 \lsun\ 
for both \lint\ and \lbol.  One source was detected in very deep, targeted 
\emph{Spitzer} 70 \um\ observations and found to have \lint\ $\sim$ 0.01 
\lsun\ and \lbol\ $\sim$ 0.2 \lsun\ (Enoch et al.~2010).  
They have all been proposed as candidate 
first hydrostatic cores, a short-lived stage between the starless and 
protostellar phases (Larson 1969).  
First cores can drive outflows, although there is still debate about the 
physical properties of such outflows and which, if any, of the sources 
have outflow properties consistent with theoretical predictions (e.g., Machida 
et al.~2008; Dunham et al.~2011; Price et al.~2012).  
Future work must be devoted to determining 
how many of these objects exist, their true evolutionary status, and the 
effects their existence will have on the protostellar luminosity distribution.

\section{Summary}\label{sec_summary}

In this paper we have studied the protostellar luminosity distribution based 
on data from two \emph{Spitzer} Legacy surveys of nearby star-forming 
regions.  We summarize our main results as follows:

\begin{enumerate}
\item Starting from a list of 2966 Young Stellar Objects identified via 
their positions in various color-color and color-magnitude diagrams, we 
identify 230 protostars in the \emph{Spitzer} c2d and Gould Belt Legacy 
surveys based on association with a dense core detected at (sub)millimeter 
wavelengths.  We compile as complete SEDs as possible for all 230 sources, and 
use these SEDs to calculate \lbol\ and \tbol.
\item The protostellar luminosity distribution extends over three orders of 
magnitude, from 0.01 \lsun\ -- 69 \lsun, and has a mean and median of 4.3 
\lsun\ and 1.3 \lsun, respectively.  Several dimensionless quantities 
characterizing the shape of the distribution are also calculated and tabulated.
\item The luminosity distributions are generally similar for Class 0 and Class 
I sources, with mean (median) values of 4.5 \lsun\ and 3.8 \lsun\ (1.4 \lsun\ 
and 1.0 \lsun) for the Class 0 and I sources, respectively.  The only 
difference is an excess of low luminosity Class I sources compared 
to the Class 0 population: 36\% of the Class I sources have \lbol\ $< 0.5$ 
\lsun\ compared to only 20\% for the Class 0 population.  A K-S test 
confirms that this difference is statistically significant.
\item Our derived luminosity distribution is similar to that obtained by 
Evans et al.~(2009) from the c2d data, except with better statistics.  The 
most significant change is that we have added additional data to improve 
the accuracy of the \lbol\ measurement for sources saturated at 24 \um\ 
with \emph{Spitzer}, reducing \lbol\ by factors of $\sim2-3$ for 14 of the 
highest luminosity sources.  This improvement is responsible for most 
of the decrease in the mean \lbol\ from 5.3 \lsun\ in Evans et al.~(2009) 
to 4.3 \lsun\ in this study.
\item Our derived luminosity distribution is significantly different from that 
of Kryukova et al.~(2012), who find a strong excess of sources at \lbol\ 
$\leq 1.0$ \lsun\ compared to our results.  Some of this discrepancy can be 
explained by the fact that we have modified the SEDs used to calculate \lbol, 
requiring a modification to the Kryukova et al.~relationship used to calculate 
\lbol, but some is also due to source selection.  By not requiring a 
(sub)millimeter detection they identify nearly 40\% more protostars than we do 
in the same clouds, most of which are located at low \lbol.  Future work is 
needed to better characterize the completeness of our sample and the 
reliability of the Kryukova et al.~sample.
\item 100 out of 230 protostars (43\%) lack any available data in the 
far-infrared 
and submillimeter.  The calculated \lbol\ for these sources underestimates the 
true \lbol\ by a factor of 2.5 on average, and up to factors of $8-10$ in 
extreme cases.  Including far-infrared and submillimeter data for 
these sources once they become available from the \emph{Herschel} Gould Belt 
survey will likely increase both the mean and median \lbol\ by 35\% -- 40\%.
\item The conclusions of several recent theoretical studies of the protostellar 
accretion process that compare to the Evans et al.~(2009) c2d luminosity 
distribution remain valid since our results are not substantially different 
from the c2d results.  As these studies 
demonstrate that there is more than one plausible accretion scenario that 
can match observations, we have emphasized that future theoretical work 
is needed to better decipher the implications of protostellar luminosities 
for the underlying mass accretion process.
\end{enumerate}

We believe that the results presented here are the most complete and reliable 
census of protostellar luminosities assembled to date.  Nevertheless, 
as outlined above in \S \ref{sec_future}, several avenues of future work must 
be pursued to better define the true completeness and reliability of the 
protostellar sample, more accurately measure \lbol\ for each source, and better 
understand the full extent and shape of the low end of the luminosity 
distribution.

\acknowledgments
We thank Erin Kryukova, Tom Megeath, Chris McKee, Stella Offner, and Phil Myers 
for reading a draft in advance of publication and providing helpful comments.  
This work is based primarily on observations obtained with the \emph{Spitzer 
Space Telescope}, operated by the Jet Propulsion Laboratory, California 
Institute of Technology.  It also uses data products from the Two Micron All 
Sky Survey, which is a joint project of the University of Massachusetts and 
the Infrared Processing and Analysis Center/California Institute of 
Technology, funded by the National Aeronautics and Space Administration and 
the National Science Foundation.  These data were provided by the NASA/IPAC 
Infrared Science Archive, which is operated by the Jet Propulsion Laboratory, 
California Institute of Technology, under contract with NASA.  This research 
has made use of NASA's Astrophysics Data System (ADS) Abstract Service, and 
the SIMBAD database, operated at CDS, Strasbourg, France.  
M.M.D.~acknowledges the support of a NASA/JPL \emph{Herschel} OT1 
grant.  Additional support was provided by the NSF through grant AST-0845619 
to H.G.A.  N.J.E. acknowledges support from the NSF through grant AST-1109116.

\clearpage

\LongTables
%\begin{landscape}
\begin{deluxetable}{llcrrrrrrrc}
\tabletypesize{\scriptsize}
\tablewidth{0pt}
\tablecaption{\label{tabysoprops}List of Protostars and Basic Properties}
\tablehead{
                &                 & \colhead{\emph{Spitzer}}      &         & \multicolumn{3}{c}{Observed}  & \multicolumn{3}{c}{Extinction Corrected}  &   \\
                &                 & \colhead{Source Name}  &         &          & \colhead{\tbol} & \colhead{\lbol}    &          & \colhead{\tbol$^{\prime}$} & \colhead{\lbol$^{\prime}$} &   \\
\colhead{Index} & \colhead{Cloud} & \colhead{(SSTc2d or SSTgb $+$)} & \colhead{A$_{\rm V}$\tablenotemark{a}} & \colhead{$\alpha$} & \colhead{(K)}    & \colhead{(\lsun)}   & \colhead{$\alpha^{\prime}$} & \colhead{(K)}    & \colhead{(\lsun)} & \colhead{FIR/SMM\tablenotemark{b}}
}
\startdata
  1   &             Aquila   &   J1829053$-$014157   &   12.4   &      1.14   &     80   &   2.7   &      0.96   &     96   &   2.9   & N  \\
  2   &             Aquila   &   J1829234$-$013856   &   12.4   &      1.39   &    250   &   0.84   &      1.08   &    320   &   1.2   & N  \\
  3   &             Aquila   &   J1829381$-$015101   &   12.4   &      1.30   &    240   &   0.33   &      1.09   &    260   &   0.46   & N  \\
  4   &             Aquila   &   J1829387$-$015100   &   12.4   &      0.82   &     84   &   3.0   &      0.65   &    110   &   3.3   & N  \\
  5   &             Aquila   &   J1829419$-$015012   &   12.4   &      0.65   &    110   &   0.067   &      0.46   &    150   &   0.078   & N  \\
  6   &             Aquila   &   J1829433$-$015652   &   12.4   &      0.81   &    310   &   0.67   &      0.94   &    380   &   0.99   & N  \\
  7   &             Aquila   &   J1829470$-$015548   &   12.4   &      0.60   &    220   &   0.85   &      0.29   &    330   &   1.1   & N  \\
  8   &             Aquila   &   J1829595$-$020106   &   12.4   &      0.74   &    180   &   0.037   &      0.53   &    230   &   0.049   & N  \\
  9   &             Aquila   &   J1830011$-$020609   &   12.4   &      0.76   &    110   &   0.85   &      0.34   &    160   &   0.98   & N  \\
 10   &             Aquila   &   J1830025$-$020258   &   12.4   &      0.48   &    390   &   1.2   &      0.36   &    470   &   1.9   & N  \\
 11   &             Aquila   &   J1830175$-$020958   &   12.4   &      1.32   &    130   &   0.24   &      1.11   &    150   &   0.30   & N  \\
 12   &             Aquila   &   J1830246$-$015411   &   12.4   &      0.21   &    400   &   0.73   &      0.12   &    590   &   1.1   & N  \\
 13   &             Aquila   &   J1830259$-$021043   &   12.4   &      1.38   &     95   &   3.5   &      1.20   &    110   &   4.0   & N  \\
 14   &             Aquila   &   J1830293$-$015643   &   12.4   &      0.21   &    480   &   1.5   &     $-$0.06   &    710   &   2.6   & N  \\
 15   &             Aquila   &   J1830469$-$015646   &   12.4   &      1.18   &    120   &   0.16   &      0.87   &    140   &   0.20   & N  \\
 16   &             Aquila   &   J1830487$-$015602   &   12.4   &      0.98   &    190   &   0.28   &      0.80   &    250   &   0.36   & N  \\
 17   &             Aquila   &   J1831522$-$020126   &   12.4   &      1.56   &    160   &   0.12   &      1.35   &    170   &   0.16   & N  \\
 18   &             Aquila   &   J1832132$-$015730   &   12.4   &      0.93   &     57   &   0.28   &      0.73   &     59   &   0.29   & N  \\
 19   &             Auriga/California   &   J0410416+380805   &   10.0   &     $-$0.32   &     50   &   7.0   &     $-$0.88   &     98   &   7.3   & Y  \\
 20   &             Auriga/California   &   J0430036+351420   &    7.5   &      0.75   &     98   &   0.75   &      0.80   &    140   &   0.80   & N  \\
 21   &             Auriga/California   &   J0430082+351410   &    7.5   &      0.46   &    130   &   0.81   &      0.44   &    180   &   0.90   & N  \\
 22   &             Auriga/California   &   J0430145+351332   &   23.4   &     $-$0.39   &    720   &   0.94   &     $-$1.26   &   1500   &   4.7   & N  \\
 23   &            Cepheus   &   J2035463+675302   &    5.4   &      1.00   &     50   &   1.4   &      0.56   &     53   &   1.4   & Y  \\
 24   &            Cepheus   &   J2036198+675631   &    6.5   &     $-$0.50   &    760   &   2.2   &     $-$0.64   &   2500   &   4.9   & Y  \\
 25   &            Cepheus   &   J2040567+672305   &    5.4   &      0.81   &    150   &   0.11   &      0.30   &    200   &   0.12   & N  \\
 26   &            Cepheus   &   J2057130+773543   &    5.4   &      0.10   &    360   &   2.0   &     $-$0.04   &    530   &   2.4   & Y  \\
 27   &            Cepheus   &   J2100207+681316   &    5.4   &      1.60   &     89   &   0.90   &      1.36   &    100   &   0.95   & Y  \\
 28   &            Cepheus   &   J2100221+681258   &    5.4   &      0.95   &     87   &   0.83   &      0.52   &    100   &   0.86   & Y  \\
 29   &            Cepheus   &   J2101328+681120   &    5.4   &      1.00   &     21   &   3.1   &      0.94   &     26   &   3.1   & Y  \\
 30   &            Cepheus   &   J2102212+675420   &    5.4   &      1.10   &    180   &   0.41   &      0.64   &    210   &   0.44   & Y  \\
 31   &            Cepheus   &   J2102273+675418   &    5.4   &     $-$0.29   &     36   &   0.32   &     $-$0.64   &     47   &   0.33   & Y  \\
 32   &            Cepheus   &   J2228030+690116   &    5.4   &      0.84   &    160   &   2.1   &      0.40   &    180   &   2.3   & Y  \\
 33   &            Cepheus   &   J2228074+690038   &    5.4   &      1.00   &     42   &   0.94   &      0.90   &     43   &   0.95   & Y  \\
 34   &            Cepheus   &   J2229333+751316   &    5.4   &      0.12   &    210   &   0.086   &      0.10   &    240   &   0.096   & N  \\
 35   &            Cepheus   &   J2229594+751403   &    5.4   &      0.49   &    270   &   0.32   &      0.16   &    330   &   0.36   & N  \\
 36   &            Cepheus   &   J2230318+751409   &    5.4   &      0.71   &     35   &   0.43   &      0.64   &     36   &   0.43   & Y  \\
 37   &            Cepheus   &   J2231056+751337   &    5.1   &     $-$0.51   &     30   &   0.30   &     $-$0.77   &     31   &   0.30   & Y  \\
 38   &            Cepheus   &   J2238428+751136   &    5.4   &      1.20   &     72   &   1.8   &      0.89   &     78   &   1.8   & Y  \\
 39   &            Cepheus   &   J2238469+751133   &    5.4   &      1.60   &    110   &   6.2   &      1.94   &    120   &   6.6   & Y  \\
 40   &            Cepheus   &   J2238530+751123   &    5.4   &      0.97   &     90   &   1.4   &      0.44   &    150   &   1.5   & Y  \\
 41   &            Cepheus   &   J2039062+680215   &    5.4   &      0.94   &     35   &   4.0   &      0.79   &     35   &   4.1   & Y  \\
 42   &            Cepheus   &   J2045539+675738   &    5.4   &   \nodata   &    340   &   48    &   \nodata   &    450   &   62    & Y  \\
 43   &            Cepheus   &   J2235234+751707   &    5.4   &      2.40   &    230   &   8.7   &      3.69   &    280   &   9.5   & Y  \\
 44   &       Chamaeleon I   &   J1104227$-$771808   &    6.6   &      0.61   &    350   &   0.086   &      0.27   &    510   &   0.11   & N  \\
 45   &       Chamaeleon I   &   J1106464$-$772232   &    6.6   &      1.23   &     66   &   0.69   &      1.11   &     74   &   0.72   & Y  \\
 46   &       Chamaeleon I   &   J1106580$-$772248   &    6.6   &      0.51   &    170   &   0.15   &      0.92   &    210   &   0.17   & N  \\
 47   &       Chamaeleon I   &   J1107161$-$772306   &    6.6   &     $-$0.10   &    490   &   0.076   &     $-$0.03   &    590   &   0.099   & N  \\
 48   &       Chamaeleon I   &   J1107213$-$772211   &    6.6   &     $-$0.08   &    650   &   0.089   &     $-$0.45   &    820   &   0.13   & N  \\
 49   &       Chamaeleon I   &   J1107435$-$773941   &    5.9   &     $-$1.05   &   1400   &   0.45   &     $-$1.43   &   1700   &   0.85   & N  \\
 50   &       Chamaeleon I   &   J1108029$-$773842   &    6.6   &     $-$0.10   &    710   &   0.97   &     $-$0.39   &    900   &   1.4   & N  \\
 51   &       Chamaeleon I   &   J1109285$-$763328   &    6.6   &      1.17   &    260   &   0.90   &      1.30   &    300   &   1.1   & Y  \\
 52   &       Chamaeleon I   &   J1109461$-$763446   &    5.9   &     $-$0.41   &    720   &   0.18   &     $-$1.13   &   1100   &   0.26   & N  \\
 53   &       Chamaeleon I   &   J1109472$-$772629   &    9.2   &     $-$0.81   &   1100   &   0.092   &     $-$1.25   &   1500   &   0.21   & N  \\
 54   &       Chamaeleon I   &   J1110033$-$763311   &    6.6   &      0.32   &    270   &   0.018   &     $-$0.18   &    430   &   0.022   & N  \\
 55   &       Chamaeleon I   &   J1110113$-$763529   &    8.0   &     $-$0.44   &   1100   &   0.37   &     $-$1.24   &   1600   &   0.80   & N  \\
 56   &       Chamaeleon I   &   J1111107$-$764157   &    6.6   &      0.09   &    330   &   0.0076   &     $-$0.20   &    650   &   0.0095   & N  \\
 57   &      Chamaeleon II   &   J1253172$-$770710   &   10.5   &     $-$0.72   &    660   &   30    &     $-$0.63   &   1500   &   63    & N  \\
 58   &      Chamaeleon II   &   J1253428$-$771511   &    4.0   &      0.65   &    130   &   0.43   &      0.32   &    160   &   0.45   & Y  \\
 59   &      Chamaeleon II   &   J1259065$-$770739   &    4.0   &      0.68   &    230   &   1.7   &      1.12   &    260   &   1.8   & Y  \\
 60   &   Corona Australis   &   J1901480$-$365722   &    7.9   &      0.78   &     93   &   4.4   &      1.03   &    130   &   4.7   & Y  \\
 61   &   Corona Australis   &   J1901484$-$365714   &    7.9   &      1.41   &     17   &   3.7   &      1.20   &     23   &   3.7   & Y  \\
 62   &   Corona Australis   &   J1901537$-$370033   &    1.5   &     $-$1.09   &     19   &   1.4   &     $-$1.29   &     23   &   1.4   & Y  \\
 63   &   Corona Australis   &   J1901585$-$365708   &    7.9   &      0.88   &     13   &   6.9   &      0.72   &     15   &   7.0   & Y  \\
 64   &   Corona Australis   &   J1902586$-$370735   &    7.9   &      1.66   &     61   &   1.2   &      1.48   &     66   &   1.3   & Y  \\
 65   &   Corona Australis   &   J1901086$-$365720   &    7.9   &     $-$0.80   &   1000   &   3.5   &     $-$1.22   &   1500   &   6.9   & Y  \\
 66   &   Corona Australis   &   J1903068$-$371249   &    7.9   &      0.36   &    460   &   8.8   &     $-$0.21   &    610   &   13    & Y  \\
 67   &   Corona Australis   &   J1901506$-$365809   &    7.9   &      0.92   &    210   &   15    &      1.01   &    270   &   17    & Y  \\
 68   &   Corona Australis   &   J1901415$-$365831   &    7.9   &      0.75   &    270   &   7.1   &      0.27   &    390   &   8.5   & Y  \\
 69   &   Corona Australis   &   J1901553$-$365721   &    7.9   &      2.64   &    260   &   0.44   &      2.47   &    310   &   0.53   & Y  \\
 70   &   Corona Australis   &   J1901564$-$365728   &    7.9   &      2.78   &    200   &   1.4   &      2.56   &    210   &   1.8   & Y  \\
 71   &             IC5146   &   J2145585+473601   &    3.6   &      0.82   &    150   &   7.4   &      1.11   &    170   &   7.8   & Y  \\
 72   &             IC5146   &   J2147227+473214   &    3.6   &      0.74   &     86   &   35    &      0.63   &     90   &   36    & Y  \\
 73   &              Lupus   &   J1539277$-$344617   &    1.0   &     $-$0.84   &   2700   &   0.94   &     $-$1.20   &   3300   &   1.3   & N  \\
 74   &              Lupus   &   J1539282$-$344618   &    2.0   &     $-$0.84   &   2100   &   0.24   &     $-$1.11   &   2600   &   0.36   & N  \\
 75   &              Lupus   &   J1607100$-$391103   &    3.0   &     $-$1.04   &   2100   &   0.69   &     $-$1.23   &   3700   &   1.6   & N  \\
 76   &              Lupus   &   J1608217$-$390421   &    1.0   &     $-$1.13   &   2600   &   0.21   &     $-$1.46   &   3100   &   0.28   & N  \\
 77   &              Lupus   &   J1608224$-$390446   &    0.0   &     $-$0.48   &   1900   &   1.7   &     $-$0.63   &   1900   &   1.7   & N  \\
 78   &              Lupus   &   J1609180$-$390453   &    2.9   &      1.10   &     39   &   0.41   &      1.14   &     39   &   0.41   & Y  \\
 79   &          Ophiuchus   &   J1625381$-$242236   &   13.2   &     $-$0.79   &   1100   &   0.11   &     $-$1.22   &   1600   &   0.36   & N  \\
 80   &          Ophiuchus   &   J1625561$-$242048   &    4.5   &     $-$0.60   &   1100   &   0.89   &     $-$0.88   &   1400   &   1.3   & N  \\
 81   &          Ophiuchus   &   J1626103$-$242054   &   27.0   &     $-$0.46   &    290   &   1.0   &     $-$1.43   &   1300   &   3.7   & Y  \\
 82   &          Ophiuchus   &   J1626146$-$242507   &    0.0   &   \nodata   &      7   &   0.034   &   \nodata   &      7   &   0.034   & N  \\
 83   &          Ophiuchus   &   J1626188$-$242819   &   19.7   &     $-$0.73   &    990   &   0.50   &     $-$1.28   &   1600   &   2.5   & N  \\
 84   &          Ophiuchus   &   J1626213$-$242304   &    9.8   &      1.46   &    210   &   8.6   &      1.14   &    250   &   11    & Y  \\
 85   &          Ophiuchus   &   J1626236$-$244314   &    4.0   &     $-$1.12   &   1500   &   0.48   &     $-$1.34   &   1700   &   0.73   & Y  \\
 86   &          Ophiuchus   &   J1626240$-$241613   &   13.3   &     $-$0.71   &    980   &   1.9   &     $-$1.09   &   1500   &   5.5   & N  \\
 87   &          Ophiuchus   &   J1626254$-$242301   &    9.8   &      0.87   &    140   &   0.010   &      0.60   &    200   &   0.12   & N  \\
 88   &          Ophiuchus   &   J1626256$-$242428   &    9.8   &      1.65   &     72   &   0.038   &      1.44   &     84   &   0.043   & N  \\
 89   &          Ophiuchus   &   J1626404$-$242714   &    9.8   &      0.45   &    380   &   0.064   &      0.32   &    470   &   0.092   & N  \\
 90   &          Ophiuchus   &   J1626441$-$243448   &    9.8   &      2.49   &    330   &   0.98   &      2.50   &    380   &   1.4   & N  \\
 91   &          Ophiuchus   &   J1626450$-$242307   &   18.5   &     $-$0.64   &    820   &   0.29   &     $-$1.20   &   1500   &   1.2   & N  \\
 92   &          Ophiuchus   &   J1626484$-$242838   &    9.8   &      0.02   &    440   &   0.12   &     $-$0.06   &    570   &   0.17   & N  \\
 93   &          Ophiuchus   &   J1626584$-$244531   &   10.7   &     $-$0.45   &    840   &   1.3   &     $-$0.70   &   1300   &   2.7   & N  \\
 94   &          Ophiuchus   &   J1627023$-$243727   &    9.8   &      1.53   &    420   &   3.3   &      0.91   &    520   &   4.8   & N  \\
 95   &          Ophiuchus   &   J1627029$-$242614   &    9.8   &     $-$0.19   &    380   &   0.036   &     $-$0.23   &    550   &   0.050   & N  \\
 96   &          Ophiuchus   &   J1627052$-$243629   &    9.8   &      1.27   &     97   &   0.16   &      1.07   &    120   &   0.18   & N  \\
 97   &          Ophiuchus   &   J1627067$-$243814   &    9.8   &      0.61   &    330   &   0.48   &      0.73   &    420   &   0.64   & N  \\
 98   &          Ophiuchus   &   J1627094$-$243718   &    9.8   &      1.69   &    370   &   13    &      1.06   &    420   &   18    & N  \\
 99   &          Ophiuchus   &   J1627158$-$243843   &   18.6   &     $-$0.70   &    220   &   0.63   &     $-$1.64   &    830   &   1.1   & N  \\
100   &          Ophiuchus   &   J1627175$-$242856   &    9.8   &      0.25   &    190   &   0.52   &      0.37   &    260   &   0.61   & N  \\
101   &          Ophiuchus   &   J1627214$-$244143   &    9.8   &     $-$0.03   &    610   &   1.1   &     $-$0.02   &    720   &   1.8   & N  \\
102   &          Ophiuchus   &   J1627218$-$242727   &    9.8   &     $-$0.05   &    180   &   0.019   &     $-$0.08   &    290   &   0.022   & N  \\
103   &          Ophiuchus   &   J1627245$-$244103   &    9.8   &      1.01   &    170   &   0.31   &      1.23   &    230   &   0.37   & N  \\
104   &          Ophiuchus   &   J1627269$-$244050   &    9.8   &      1.17   &    240   &   2.6   &      1.15   &    300   &   3.3   & N  \\
105   &          Ophiuchus   &   J1627279$-$243933   &    9.8   &      2.29   &    260   &   5.0   &      2.13   &    280   &   7.1   & N  \\
106   &          Ophiuchus   &   J1627284$-$242721   &    9.8   &     $-$0.03   &    310   &   0.48   &     $-$0.13   &    450   &   0.63   & N  \\
107   &          Ophiuchus   &   J1627301$-$242743   &    9.8   &     $-$0.12   &    500   &   0.97   &     $-$0.02   &    620   &   1.5   & N  \\
108   &          Ophiuchus   &   J1627372$-$244237   &    9.8   &      0.13   &    460   &   0.12   &      0.15   &    560   &   0.18   & N  \\
109   &          Ophiuchus   &   J1627398$-$244315   &    9.8   &     $-$0.15   &    570   &   0.72   &     $-$0.15   &    690   &   1.1   & N  \\
110   &          Ophiuchus   &   J1628216$-$243623   &    9.8   &      1.23   &     33   &   0.24   &      0.96   &     36   &   0.24   & Y  \\
111   &          Ophiuchus   &   J1628578$-$244054   &    9.8   &      0.67   &    320   &   0.027   &      0.58   &    430   &   0.037   & N  \\
112   &          Ophiuchus   &   J1631356$-$240129   &    9.8   &      0.14   &    270   &   1.6   &      0.14   &    390   &   2.0   & Y  \\
113   &          Ophiuchus   &   J1631367$-$240419   &    9.8   &     $-$0.27   &     74   &   0.17   &     $-$0.22   &    160   &   0.19   & Y  \\
114   &          Ophiuchus   &   J1631437$-$245524   &    9.8   &      0.23   &    520   &   0.26   &      0.19   &    690   &   0.40   & N  \\
115   &          Ophiuchus   &   J1631520$-$245726   &    9.8   &      0.82   &    120   &   0.0082   &      0.61   &    150   &   0.0095   & N  \\
116   &          Ophiuchus   &   J1631524$-$245536   &    9.8   &      1.07   &    260   &   0.11   &      0.87   &    330   &   0.15   & N  \\
117   &          Ophiuchus   &   J1632009$-$245642   &    9.8   &      1.39   &    140   &   2.5   &      1.36   &    180   &   2.8   & Y  \\
118   &          Ophiuchus   &   J1632226$-$242831   &    9.8   &      5.03   &     45   &   8.5   &      4.87   &     45   &   8.8   & Y  \\
119   &          Ophiuchus   &   J1633556$-$244205   &    4.6   &     $-$1.22   &   1500   &   0.17   &     $-$1.31   &   1800   &   0.29   & N  \\
120   &           Ophiuchus North   &   J1646582$-$093519   &    5.6   &      0.66   &    230   &   0.44   &      0.61   &    280   &   0.50   & Y  \\
121   &           Ophiuchus North   &   J1648456$-$141636   &    5.6   &     $-$0.97   &   1400   &   1.2   &     $-$1.39   &   1700   &   2.2   & N  \\
122   &           Ophiuchus North   &   J1657196$-$160923   &    5.7   &      2.40   &     38   &   0.80   &      2.57   &     39   &   0.82   & Y  \\
123   &            Perseus   &   J0325223+304513   &    5.9   &      2.34   &     52   &   2.0   &      2.20   &     53   &   2.1   & Y  \\
124   &            Perseus   &   J0325362+304515   &    5.9   &      1.59   &     12   &   8.5   &      1.49   &     12   &   8.5   & Y  \\
125   &            Perseus   &   J0325364+304522   &    5.9   &      2.62   &     66   &   4.9   &      2.55   &     70   &   5.2   & Y  \\
126   &            Perseus   &   J0325388+304406   &    5.9   &      2.16   &     47   &   6.9   &      2.04   &     48   &   7.0   & Y  \\
127   &            Perseus   &   J0325391+304358   &    5.9   &      2.36   &    160   &   0.69   &      2.02   &    170   &   0.78   & N  \\
128   &            Perseus   &   J0326374+301528   &    5.9   &      1.09   &     64   &   0.91   &      1.02   &     73   &   0.94   & Y  \\
129   &            Perseus   &   J0327382+301358   &    5.9   &     $-$0.19   &    260   &   0.78   &     $-$0.40   &    360   &   0.89   & Y  \\
130   &            Perseus   &   J0327390+301303   &    5.9   &      2.68   &     62   &   3.5   &      2.45   &     65   &   3.6   & Y  \\
131   &            Perseus   &   J0327432+301228   &    5.9   &      2.39   &     54   &   1.7   &      2.23   &     57   &   1.7   & Y  \\
132   &            Perseus   &   J0327476+301204   &    5.9   &     $-$0.09   &    740   &   2.5   &     $-$0.25   &    950   &   3.6   & Y  \\
133   &            Perseus   &   J0328003+300801   &    5.9   &      0.95   &    230   &   0.25   &      0.99   &    280   &   0.29   & Y  \\
134   &            Perseus   &   J0328325+311105   &    5.9   &      0.78   &     52   &   0.26   &      0.50   &     74   &   0.27   & Y  \\
135   &            Perseus   &   J0328344+310051   &    5.9   &      0.83   &    240   &   1.1   &      0.88   &    290   &   1.3   & Y  \\
136   &            Perseus   &   J0328345+310705   &    5.9   &      0.54   &    150   &   0.12   &      0.32   &    170   &   0.13   & Y  \\
137   &            Perseus   &   J0328350+302009   &    5.9   &      0.15   &     49   &   0.36   &     $-$0.22   &     89   &   0.36   & Y  \\
138   &            Perseus   &   J0328370+311330   &    5.9   &      2.35   &    100   &   9.5   &      1.94   &    110   &   10    & Y  \\
139   &            Perseus   &   J0328391+310601   &    5.9   &      1.68   &     28   &   0.23   &      1.55   &     29   &   0.23   & Y  \\
140   &            Perseus   &   J0328397+311731   &    5.9   &      0.57   &    250   &   0.18   &      0.61   &    300   &   0.21   & Y  \\
141   &            Perseus   &   J0328406+311756   &    5.9   &      1.02   &     12   &   0.58   &      0.95   &     12   &   0.58   & Y  \\
142   &            Perseus   &   J0328432+311732   &    5.9   &      0.36   &    490   &   1.6   &      0.25   &    640   &   2.0   & Y  \\
143   &            Perseus   &   J0328453+310541   &    5.9   &      1.11   &     62   &   0.41   &      1.08   &     72   &   0.43   & Y  \\
144   &            Perseus   &   J0328555+311436   &    5.9   &      3.03   &     54   &   22    &      2.37   &     55   &   22    & Y  \\
145   &            Perseus   &   J0328563+312227   &    5.9   &     $-$0.14   &    440   &   0.15   &     $-$0.40   &    620   &   0.19   & N  \\
146   &            Perseus   &   J0328573+311415   &    5.9   &      1.60   &    100   &   5.3   &      1.40   &    110   &   5.7   & Y  \\
147   &            Perseus   &   J0328584+312217   &    5.9   &      0.83   &    240   &   0.96   &      0.83   &    280   &   1.1   & Y  \\
148   &            Perseus   &   J0328593+311548   &    5.9   &      0.06   &     10   &   3.2   &     $-$0.08   &    140   &   3.4   & Y  \\
149   &            Perseus   &   J0329005+311200   &    5.9   &      2.16   &     30   &   0.66   &      1.96   &     30   &   0.67   & Y  \\
150   &            Perseus   &   J0329015+312020   &    5.9   &      2.09   &    230   &   8.2   &      2.30   &    270   &   9.4   & Y  \\
151   &            Perseus   &   J0329033+312314   &    5.9   &      1.13   &    320   &   0.088   &      1.01   &    370   &   0.11   & N  \\
152   &            Perseus   &   J0329037+311603   &    5.9   &      1.21   &    170   &   33    &      0.85   &    220   &   37    & Y  \\
153   &            Perseus   &   J0329040+311446   &    5.9   &      1.43   &     17   &   0.68   &      1.31   &     19   &   0.68   & Y  \\
154   &            Perseus   &   J0329077+312157   &    5.9   &      2.18   &    230   &   18    &      2.31   &    260   &   21    & Y  \\
155   &            Perseus   &   J0329104+311331   &    5.9   &      2.58   &     31   &   7.9   &      2.48   &     31   &   8.0   & Y  \\
156   &            Perseus   &   J0329106+311820   &    5.9   &      1.95   &     56   &   3.2   &      1.84   &     58   &   3.3   & Y  \\
157   &            Perseus   &   J0329112+311831   &    5.9   &      1.94   &     29   &   1.2   &      1.77   &     32   &   1.3   & Y  \\
158   &            Perseus   &   J0329120+311305   &    5.9   &      0.98   &     25   &   4.2   &      0.81   &     25   &   4.2   & Y  \\
159   &            Perseus   &   J0329129+311814   &    5.9   &      1.05   &    240   &   0.100   &      1.58   &    270   &   1.1   & N  \\
160   &            Perseus   &   J0329135+311358   &    5.9   &      2.41   &     35   &   0.84   &      2.50   &     36   &   0.85   & Y  \\
161   &            Perseus   &   J0329171+312746   &    5.9   &      1.75   &     32   &   0.65   &      1.70   &     34   &   0.66   & Y  \\
162   &            Perseus   &   J0329182+312319   &    5.9   &      1.26   &     21   &   0.54   &      1.11   &     24   &   0.54   & Y  \\
163   &            Perseus   &   J0329187+312325   &    5.9   &     $-$0.23   &    190   &   1.9   &     $-$0.67   &    410   &   2.1   & Y  \\
164   &            Perseus   &   J0329200+312407   &    5.9   &      0.42   &     75   &   1.5   &      0.34   &     10   &   1.6   & Y  \\
165   &            Perseus   &   J0329234+313329   &    5.9   &      1.51   &     60   &   0.36   &      1.51   &     64   &   0.37   & Y  \\
166   &            Perseus   &   J0329518+313906   &    5.9   &      3.44   &     39   &   0.50   &      3.40   &     40   &   0.51   & Y  \\
167   &            Perseus   &   J0330151+302349   &    5.9   &      1.70   &     93   &   1.4   &      1.57   &    100   &   1.5   & Y  \\
168   &            Perseus   &   J0330326+302626   &    5.9   &      2.08   &     34   &   0.16   &      2.72   &     35   &   0.16   & Y  \\
169   &            Perseus   &   J0331209+304530   &    5.9   &      0.98   &     32   &   1.2   &      1.48   &     32   &   1.2   & Y  \\
170   &            Perseus   &   J0332179+304947   &    5.9   &      1.07   &     25   &   1.3   &      0.87   &     26   &   1.3   & Y  \\
171   &            Perseus   &   J0332291+310240   &    5.9   &      0.40   &    120   &   0.52   &      0.16   &    160   &   0.56   & Y  \\
172   &            Perseus   &   J0333095+310531   &    5.9   &      1.13   &    210   &   0.042   &      0.69   &    320   &   0.049   & N  \\
173   &            Perseus   &   J0333128+312124   &    5.9   &      0.41   &    480   &   2.9   &      0.16   &    610   &   3.8   & Y  \\
174   &            Perseus   &   J0333138+312005   &    5.9   &      1.44   &     59   &   0.100   &      1.31   &     69   &   0.11   & Y  \\
175   &            Perseus   &   J0333143+310710   &    5.9   &      2.22   &     34   &   0.63   &      2.73   &     36   &   0.64   & Y  \\
176   &            Perseus   &   J0333164+310652   &    5.9   &      1.73   &     26   &   1.0   &      1.56   &     26   &   1.0   & Y  \\
177   &            Perseus   &   J0333166+310755   &    5.9   &      1.57   &    100   &   1.6   &      1.69   &    120   &   1.7   & Y  \\
178   &            Perseus   &   J0333178+310931   &    5.9   &      3.33   &     41   &   3.5   &      2.92   &     44   &   3.6   & Y  \\
179   &            Perseus   &   J0333203+310721   &    5.9   &      0.88   &     47   &   0.60   &      0.61   &     55   &   0.61   & Y  \\
180   &            Perseus   &   J0333272+310710   &    5.9   &      1.93   &     62   &   1.3   &      1.79   &     66   &   1.4   & Y  \\
181   &            Perseus   &   J0342021+314802   &    5.9   &      1.47   &    190   &   0.069   &      1.23   &    240   &   0.080   & N  \\
182   &            Perseus   &   J0343451+320358   &    5.9   &     $-$0.22   &    540   &   0.71   &      0.12   &    650   &   0.93   & N  \\
183   &            Perseus   &   J0343509+320324   &    5.9   &      1.51   &     16   &   0.68   &      1.38   &     17   &   0.69   & Y  \\
184   &            Perseus   &   J0343510+320308   &    5.9   &     $-$0.28   &     52   &   0.40   &     $-$0.30   &     57   &   0.41   & Y  \\
185   &            Perseus   &   J0343565+320052   &    5.9   &      0.50   &     23   &   1.6   &      0.55   &     23   &   1.6   & Y  \\
186   &            Perseus   &   J0343568+320304   &    5.9   &      1.37   &     22   &   1.4   &      1.83   &     23   &   1.4   & Y  \\
187   &            Perseus   &   J0343596+320154   &   19.8   &     $-$0.34   &    620   &   1.8   &     $-$0.90   &   1300   &   6.2   & Y  \\
188   &            Perseus   &   J0344024+320204   &    5.9   &      1.53   &     41   &   0.35   &      1.47   &     45   &   0.35   & Y  \\
189   &            Perseus   &   J0344129+320135   &    5.9   &      0.37   &    400   &   1.6   &      0.30   &    470   &   2.0   & N  \\
190   &            Perseus   &   J0344213+315932   &    5.9   &      0.21   &    350   &   0.26   &      0.53   &    420   &   0.32   & Y  \\
191   &            Perseus   &   J0344433+320131   &    5.9   &      0.50   &    460   &   1.2   &      0.62   &    500   &   1.5   & N  \\
192   &            Perseus   &   J0344439+320136   &    5.9   &      0.96   &     39   &   3.1   &      0.73   &     41   &   3.1   & Y  \\
193   &            Perseus   &   J0347054+324308   &    5.9   &      0.36   &    330   &   0.49   &      0.52   &    390   &   0.58   & Y  \\
194   &            Perseus   &   J0347415+325144   &    5.9   &      0.78   &    290   &   4.4   &      1.33   &    330   &   5.1   & Y  \\
195   &            Serpens   &   J1828440+005337   &    9.6   &      0.45   &    380   &   0.49   &      0.34   &    490   &   0.70   & N  \\
196   &            Serpens   &   J1828447+005125   &    9.6   &      1.07   &     97   &   0.11   &      0.85   &    120   &   0.12   & N  \\
197   &            Serpens   &   J1828449+005203   &    9.6   &      1.33   &     53   &   3.9   &      1.12   &     62   &   4.1   & Y  \\
198   &            Serpens   &   J1828512+001927   &    9.6   &      0.45   &    260   &   0.14   &      0.28   &    390   &   0.18   & N  \\
199   &            Serpens   &   J1828540+002930   &    9.6   &      1.36   &     60   &   8.3   &      1.14   &     69   &   8.8   & Y  \\
200   &            Serpens   &   J1828548+002952   &    9.6   &      1.91   &     47   &   6.3   &      1.60   &     51   &   6.6   & Y  \\
201   &            Serpens   &   J1828549+001832   &    9.6   &      0.90   &    120   &   0.12   &      0.68   &    150   &   0.14   & N  \\
202   &            Serpens   &   J1828557+002944   &    9.6   &      1.89   &     22   &   1.5   &      1.65   &     26   &   1.6   & Y  \\
203   &            Serpens   &   J1829021+003120   &    9.6   &      0.24   &     87   &   0.14   &      0.19   &    100   &   0.16   & N  \\
204   &            Serpens   &   J1829028+003009   &    9.6   &     $-$0.14   &    460   &   0.47   &     $-$0.64   &    780   &   0.70   & N  \\
205   &            Serpens   &   J1829062+003043   &    9.6   &      1.70   &     56   &   9.6   &      1.34   &     67   &   10    & Y  \\
206   &            Serpens   &   J1829067+003034   &    9.6   &      1.66   &     77   &   5.0   &      1.38   &     83   &   5.4   & N  \\
207   &            Serpens   &   J1829090+003132   &    9.6   &      2.27   &     35   &   4.1   &      2.12   &     36   &   4.2   & Y  \\
208   &            Serpens   &   J1829161+001822   &    9.6   &     $-$0.07   &    440   &   4.6   &      0.07   &    620   &   6.7   & Y  \\
209   &            Serpens   &   J1829319+011842   &    9.6   &      0.26   &    450   &   14    &     $-$0.06   &    710   &   21    & Y  \\
210   &            Serpens   &   J1829481+011644   &    9.6   &      1.37   &     29   &   14    &      1.11   &     30   &   14    & Y  \\
211   &            Serpens   &   J1829491+011619   &    9.6   &      3.80   &    130   &   11    &      3.45   &    150   &   13    & N  \\
212   &            Serpens   &   J1829496+011521   &    9.6   &      2.65   &     13   &   69    &      2.53   &     13   &   69    & Y  \\
213   &            Serpens   &   J1829511+011640   &    9.6   &      1.04   &    130   &   4.8   &      1.07   &    170   &   5.4   & Y  \\
214   &            Serpens   &   J1829522+011547   &    9.6   &      1.54   &     62   &   7.6   &      1.26   &     69   &   8.1   & Y  \\
215   &            Serpens   &   J1829525+003611   &    9.6   &      0.77   &     51   &   1.8   &      0.53   &     68   &   1.9   & Y  \\
216   &            Serpens   &   J1829528+011456   &    9.6   &      1.49   &    120   &   3.0   &      1.19   &    140   &   3.4   & N  \\
217   &            Serpens   &   J1829543+003601   &    9.6   &     $-$0.18   &     42   &   1.7   &     $-$0.34   &     59   &   1.7   & Y  \\
218   &            Serpens   &   J1829568+011446   &    9.6   &      0.30   &    330   &   16    &      0.33   &    500   &   21    & N  \\
219   &            Serpens   &   J1829575+011300   &    9.6   &      1.01   &     42   &   28    &      1.14   &     59   &   29    & Y  \\
220   &            Serpens   &   J1829577+011405   &    9.6   &      0.28   &    530   &   42    &     $-$0.08   &    700   &   64    & Y  \\
221   &            Serpens   &   J1829578+011251   &    9.6   &      0.67   &    320   &   4.5   &      0.30   &    420   &   6.2   & N  \\
222   &            Serpens   &   J1829587+011426   &    9.6   &      0.46   &    310   &   1.6   &      0.45   &    430   &   2.2   & N  \\
223   &            Serpens   &   J1829592+011401   &    9.6   &      1.03   &     41   &   7.1   &      0.87   &     44   &   7.3   & Y  \\
224   &            Serpens   &   J1829595+011159   &    9.6   &      1.15   &     87   &   14    &      1.49   &    120   &   15    & Y  \\
225   &            Serpens   &   J1829599+011311   &    9.6   &      2.57   &    100   &   6.2   &      2.22   &    120   &   7.0   & Y  \\
226   &            Serpens   &   J1830003+010944   &    9.6   &     $-$0.12   &    280   &   0.48   &     $-$0.41   &    340   &   0.61   & N  \\
227   &            Serpens   &   J1830007+011301   &    9.6   &      1.68   &     28   &   8.0   &      1.51   &     29   &   8.1   & Y  \\
228   &            Serpens   &   J1830027+011228   &    9.6   &      0.21   &    390   &   5.5   &      0.19   &    540   &   7.7   & N  \\
229   &            Serpens   &   J1830052+004104   &    9.6   &      1.30   &     95   &   0.19   &      1.04   &    100   &   0.21   & N  \\
230   &            Serpens   &   J1830057+003931   &   30.0   &     $-$0.39   &    540   &   0.18   &     $-$1.18   &   1500   &   1.1   & N  \\
\enddata 
\tablenotetext{a}{Value of A$_{\rm V}$ used for dereddening, as explained in the text.}
\tablenotetext{b}{Flag indicating that the protostar does (\emph{Y}) or does not (\emph{N}) have at least one observed photometry point in the far-infrared or submillimeter (70 \um\ $< \lambda <$ 850 \um; see \S \ref{sec_discussion_firsmm} for details).}
\end{deluxetable}
%\end{landscape}

\end{document}